\documentclass[10pt,english]{article}
\usepackage{mathptmx}
\usepackage[T1]{fontenc}
\usepackage[utf8]{luainputenc}
\usepackage{xcolor}
\usepackage{pdfcolmk}
\usepackage{babel}
\usepackage{mathrsfs}
\usepackage{amsmath}
\usepackage{amsthm}
\usepackage{amssymb}
\usepackage{graphicx}
\usepackage{setspace}
\usepackage{esint}
\PassOptionsToPackage{normalem}{ulem}
\usepackage{ulem}
\usepackage[unicode=true,pdfusetitle,
 bookmarks=false,
 breaklinks=false,pdfborder={0 0 1},backref=false,colorlinks=false]
 {hyperref}

\makeatletter

\providecolor{lyxadded}{rgb}{0,0,1}
\providecolor{lyxdeleted}{rgb}{1,0,0}

\numberwithin{equation}{section}
\numberwithin{figure}{section}
\theoremstyle{plain}
\newtheorem{thm}{\protect\theoremname}
  \theoremstyle{definition}
  \newtheorem{defn}[thm]{\protect\definitionname}
  \theoremstyle{plain}
  \newtheorem*{lem*}{\protect\lemmaname}
 \theoremstyle{definition}
 \newtheorem*{defn*}{\protect\definitionname}
  \theoremstyle{plain}
  \newtheorem{lem}[thm]{\protect\lemmaname}

\makeatother

  \providecommand{\definitionname}{Definition}
  \providecommand{\lemmaname}{Lemma}
\providecommand{\theoremname}{Theorem}

\begin{document}

\title{\noindent Replica Symmetry Breaking without Replicas}

\date{~}

\author{\noindent Simone Franchini \thanks{Goethe Mathematics Institute, 10 Robert Mayer Str, 60325 Frankfurt,
Germany}~\thanks{Sapienza Università di Roma, Piazzale Aldo Moro 1, 00185 Roma, Italy}}
\maketitle
\begin{abstract}
\noindent We introduce a mathematical framework based on simple combinatorial
arguments (Kernel Representation) that allows to deal successfully
with spin glass problems, among others. Let $\Omega^{N}$ be the space
of the configurations of an $N-$spins system, each spin having a
finite set $\Omega$ of inner states, and let $\mu:\Omega^{N}\rightarrow\left[0,1\right]$
be some probability measure. Here we give an ar\-gu\-ment to encode
$\mu$ into a kernel function $M:\left[0,1\right]^{2}\rightarrow\Omega$,
and use this notion to reinterpret the assumptions of the Replica
Symmetry Breaking ansatz (RSB) of Parisi et Al. \cite{Parisi-1,Parisi-2},
without using replicas, nor averaging on the disorder.\footnote{MSC2000: 82D30, 60F10, Keywords: Spin Glasses, Replica Symmetry Breaking,
Pure States, Parisi Formula}
\end{abstract}

\section{Introduction}

\noindent Originally introduced by Parisi \cite{Parisi-1} in his
analysis of the Sherrington-Kirkpatrick model (SK) \cite{Parisi-1,Parisi-2},
the Replica Symmetry Breaking (RSB) ansatz proved to be an extremely
valuable tool in explaining properties of disordered systems. Despite
many technical advances, worth to cite is the proof of the free energy
formula by Guerra and Talagrand \cite{Guerra,Talagarand}, some of
its fundamental features remain quite mysterious after forty years. 

A central role is played by the elusive concept of \textit{pure state}.
Despite a precise de\-fi\-nition is still lacking, it is widely
acknowledged that they must satisfy some properties. For example,
it is expected that the connected correlation functions associated
to these states vanish in the thermodynamic limit \cite{Parisi-2,Marinari}.
This imply that in some sense the mea\-sure conditioned to those
states can be described by a mean field model of some kind (see Part
III of \cite{Marinari}, updated 2014 version, for a non-rigorous
but detailed discussion of the \textit{finite volume pure states}).

Perhaps, the most striking and unconventional property is that the
pure states have be\-en pre\-dic\-ted to have a hierarchical structure,
such that the support of the overlaps is ultrametric \cite{Parisi-2}.
A considerable amount of works have been published on this argument,
that cul\-mi\-na\-ted in a proof of ultrametricity by Panchenko
\cite{Panchenko}. 

Anyway, whether the ultrametricity and other properties of the pure
states hold in some general framework, including their representation
as well defined mathematical objects, proved to be an extremely hard
task and remains an open question. 

Inspired by a remarkable series of papers by Coja-Oghlan and others,
which introdu\-ce tools from Graph Theory to study Belief Propagation
algorithms \cite{ACO-1,ACO2,ACO3,ACO4}, he\-re we present an original
framework that is based on elementary combinatorial ar\-gu\-ments,
and that allows to deal with many interesting physical systems, including
spin glasses, without using replicas, nor averaging over the disorder. 

The theory is presented both symbolically and by a graphical representation
in terms of kernel functions of the kind 
\begin{equation}
M:\,[0,1]^{2}\rightarrow\{-,+\}.
\end{equation}
This object is intended to provide a simple visual encoding for probability
measures, and it was central for us in understanding and developing
the concepts we are going to explain. 

We introduce the kernel representation in the Section \ref{sec:Kernel-representation}
along with some notation, sho\-wing how to encode a probability measure
into a kernel and bring it back, the com\-mu\-ta\-tion relations,
the transposed measure, and other basic kernel features. 

In the Section \ref{sec:Kernel-representation} we also introduce
an analogue notion of pure states, that can be ap\-plied to any distribution,
the space of these generalized states is charted by a simple par\-ti\-tion
of the spin space into disjoint subsets. This first section does not
contain com\-plex mathematical concepts, and it aims to carefully
introduce pro\-bability in kernel lan\-guage to connect with graph
theoretic arguments, and highlight the bridges that ex\-ist between
probability and graph theory. 

Then in Section \ref{sec:Kernel-filtration} we introduce more advanced
kernel methods, for example it is pos\-sible to introduce a new type
of convergence of random variables, ie conver\-gence in ``Cut Distance'',
that is stronger than weak{*} convergence and allows to manage the
kernels (then also Gibbs measures) directly in the thermodynamic limit.
In section \ref{sec:Kernel-filtration} we also give an alternative
formulation that connects with the findings presented in \cite{ACO-1,ACO2,ACO3,ACO4}
and Graph Theory in general \cite{Lovasz}. 

The readers mostly interested in the physics of the SK model may jump
these two sections in first read, and go straight to the Sec\-tion
\ref{sec:The-Ansatz}, where we apply some of the ideas pre\-sen\-ted
in Section \ref{sec:Kernel-representation} to the SK Hamiltonian
\begin{equation}
\boldsymbol{H}_{sk}\left(\sigma_{V}\right)=\frac{1}{\sqrt{N}}\sum_{i\in V}\sum_{j<i}\sigma_{i}\boldsymbol{J}_{ij}\sigma_{j}
\end{equation}
and its kernel, and develop a scheme for the propagation of properties
of the Gibbs distribution as it evolves along the cavity chain described
in \cite{Franchini}. 

The main steps of our analysis will be as follows: the first is to
define a se\-quen\-ce of SK mo\-dels of in\-creasing sizes, this
is done by partitioning the vertex set $V$ into $L$ subsets $V_{\ell}$,
marked by the index $\ell$, that we call layers. Then, we consider
their union $Q_{\ell}$ up to a certain $\ell$, with $Q_{\ell-1}\subseteq Q_{\ell}$
being a sequence of sets of size $\left|Q_{\ell}\right|=q_{\ell}N$,
such that $V_{\ell}=Q_{\ell}\setminus Q_{\ell-1}$ and $Q_{L}=V$,
see Definition \ref{11(Filtration-of-).}. 

This construction converges to the actual system, and we interpret
it as a lay\-ering scheme in which we grow the system layer by layer
up to the original size. We find the Hamiltonian sequence of these
layers in Lemma \ref{lem:(Layer-States-of}, 
\begin{equation}
\boldsymbol{H}_{sk}\left(\sigma_{V}\right)=\sum_{\ell\leq L}\boldsymbol{H}_{\ell}\left(\sigma_{Q_{\ell}}\right),
\end{equation}
where $\boldsymbol{H}_{\ell}$ is the Hamiltonian describing the layer
$V_{\ell}$, 
\begin{equation}
\boldsymbol{H}_{\ell}\left(\sigma_{Q_{\ell}}\right)=\sqrt{q_{\ell}-q_{\ell-1}}\,\boldsymbol{H}_{sk}\left(\sigma_{V_{\ell}}\right)+\sqrt{q_{\ell-1}}\sum_{i\in V_{\ell}}\sigma_{i}\boldsymbol{h}_{i}\left(\sigma_{Q_{\ell-1}}\right),
\end{equation}
the first term is the energy contribution coming from the interactions
inside the la\-yer it\-self, and is simply a smaller SK model, while
$\boldsymbol{h}_{i}$ is a cavity field only depending on the spins
of the previous layers, ie the spins of $Q_{\ell-1}$. Then we show
that, due to the me\-an field nature of the model, if the partition
is fine en\-ough the factor $\sqrt{q_{\ell}-q_{\ell-1}}$ kills the
contribution from the smaller SK, and the thermodynamics is dominated
by the interface, ie the interaction term $\sigma_{V_{\ell}}\cdot\boldsymbol{h}_{V_{\ell}}$
between the newly added layer and the rest of the system up to that
point of the sequence, see Lemma \ref{lem:(IO-model)-For}. 

The interface is simply a collection of spins coupled to an external
random field that comes from the previous layers, and can be solved
exactly, for example by the same tech\-niques developed in \cite{BCP theorem,BCP 2,Bauke Frnz Mertens}
for the Number Partitioning Problem. In Section \ref{sec:Thermodynamics-of-the}
we show that the thermodynamics of the layer is equivalent to a Random
Energy Model of the Derrida type \cite{Bovier-Kurkova-1} in the limit
of very low tem\-pe\-ra\-tu\-re, that is Lemma \ref{lem:(Random-Energy-Model-1}. 

Finally, in Section \ref{sec:ROSt-variables-and} we give an explicit
application by combining the Cavity Me\-thod of \cite{UltraBolt,ASS,Bolt,Franchini}
with Lemma \ref{11(Filtration-of-).} to obtain a constructive derivation
of the cavity variables, and compute the Parisi functional in a simple
way. 

\pagebreak{}

\section{Kernel representation\label{sec:Kernel-representation}}

Before entering in the core of the discussion some preliminaries are
mandatory in order to explain the notation and justify our later arguments.
In particular, we describe how and why to encode a finite spin system
into a kernel function. For this paper we indicate by $I\left(A\right)$
the in\-dicator function of the event $A$, that is $I\left(A\right)=1$
if $A$ is verified and is zero otherwise. Also, given two ordered
sets $A$ and $B$ we use the notation $A\otimes B$ for the tensor
product and just $AB$ for the Cartesian product (ie same for number
multiplication). The Hadamard product is denoted by the $\circ$ symbol. 

Consider a random spin system $\boldsymbol{\sigma}_{V}$ of $N$ spins,
distributed according to some law $\mu\left(\sigma_{V}\right)$, and
imagine to perform a sequence of independent measurements of such
system. Formally, let $V=\left\{ 1,2,\,...\,,N\right\} $ be a set
of $N$ vertices and put a spin $\sigma_{i}\in\Omega$ of inner states
$\Omega=\{+,-\}$ on each vertex, we denote by 
\begin{equation}
\sigma_{V}=\left\{ \sigma_{i}\in\Omega:\,i\in V\right\} 
\end{equation}
the generic magnetization state. Due to the finiteness of the spin
number there is only a finite set of possible outcomes, in fact, each
measurement will give as result some element of $\Omega^{V}$ product
space of the elementary spin spaces $\Omega$. 

The first important observation is that if the mea\-surements are
independent the order in which the states are observed cannot contain
information of the underlying law, then we are free to regroup them
to our convenience. Let order the states of $\Omega^{V}$ by some
index $\alpha:\,\Omega^{V}\rightarrow S$, where $S=\{1,2,\,...\,,2^{|V|}\}$
is the span of the index. The set $\Omega^{V}$ is then rewritten
as follows 
\begin{equation}
\Omega^{V}=\{\tau_{V}^{\alpha}:\,\alpha\in S\},\,\,\tau_{V}^{\alpha}=\{\tau_{i}^{\alpha}\in\Omega:\,i\in V\}
\end{equation}
with each state $\tau_{V}^{\alpha}$ being uniquely identified by
$\alpha$, ie $\tau_{V}^{\alpha}\neq\tau_{V}^{\gamma}$ if $\alpha\neq\gamma$. 

Since for finite $V$ also $\Omega^{V}$ has a finite number of states,
for a large number of mea\-su\-re\-ments the relative frequencies
of the states $\tau_{V}^{\alpha}$, that are rational numbers, ap\-pro\-xi\-ma\-te
the probabilities $\mu\left(\tau_{V}^{\alpha}\right)\in\left[0,1\right]$
that are associated to the occurrence of a given state $\alpha$.
Arranging them into vectors
\begin{equation}
\mu=\{\mu\left(\tau_{V}^{\alpha}\right)\in\left[0,1\right]:\,\alpha\in S\},
\end{equation}
we can also write a simple representation for the set of measures
on $\Omega^{V}$ 
\begin{equation}
\mathcal{P}\left(\Omega^{V}\right)={\textstyle \{\mu\in\left[0,1\right]^{V}:\,{\textstyle \sum_{\alpha}}\,\mu\left(\tau_{V}^{\alpha}\right)=1\}}.
\end{equation}
Is easy to verify that the measure (probability mass function) can
be reconstructed from the vector $\mu$. Explicitly, we can write
the measure $\mu:\Omega^{V}\rightarrow[0,1]$ and and its average
applied to some test function $f:\Omega^{V}\rightarrow\mathbb{R}$
as follows 
\begin{equation}
\mu\left(\sigma_{V}\right)=\sum_{\alpha\in S}\mu\left(\tau_{V}^{\alpha}\right)\prod_{i\in V}{\textstyle \left(\frac{1+\tau_{i}^{\alpha}\,\sigma_{i}}{2}\right)},\ \ \langle f\left(\boldsymbol{\sigma}_{V}\right)\rangle_{\mu}=\sum_{\alpha\in S}\mu\left(\tau_{V}^{\alpha}\right)f\left(\tau_{V}^{\alpha}\right).\label{eq:definitionsssss}
\end{equation}
This will be our preferential notation, and we will also use a dedicated
symbol for the uniform measure $\nu^{\alpha}=1/|S|=1/2^{N}$ and call
it \textit{support measure} (see upper kernel of Figure \ref{fig3.1})
\begin{equation}
\nu\left(\sigma_{V}\right)=\frac{1}{2^{N}}\sum_{\alpha\in S}\prod_{i\in V}{\textstyle \left(\frac{1+\tau_{i}^{\alpha}\,\sigma_{i}}{2}\right)},\ \ \langle f\left(\boldsymbol{\sigma}_{V}\right)\rangle_{\nu}=\frac{1}{2^{N}}\sum_{\alpha\in S}f\left(\tau_{V}^{\alpha}\right).
\end{equation}

We can now introduce a powerful graphical tool to represent $(\mu,\Omega^{V})$,
that simply consists in rearranging the states into an array. In the
following we show how to encode the probability pair $(\mu,\Omega^{V})$
into a two dimensional function. 
\begin{defn}
\textit{\label{1(Magnetization-Kernel)-Let}(Magnetization Kernel)
Let $\mu\in\mathcal{P}\left(\Omega^{V}\right)$, then, the Magnetization
Kernel $M:\left[0,1\right]^{2}\rightarrow\Omega$ associated to $\mu$
is the step function 
\begin{equation}
M_{x}^{y}=\sum_{\alpha\in S}\sum_{i\in V}\,\tau_{i}^{\alpha}\,I\left(\,x\in\left(x{}_{i-1},x_{i}\right],\,y\in\left(y_{\alpha-1},y_{\alpha}\right]\,\right)\label{eq:magkernel}
\end{equation}
with $I(A)$ indicator function of the event $A$. The sizes of the
intervals are 
\begin{equation}
x_{i}-x_{i-1}=1/N,\ y_{\alpha}-y_{\alpha-1}=\mu\left(\tau_{V}^{\alpha}\right).\label{eq:compactification}
\end{equation}
An explicit example is given in Figures \ref{fig3.1} and \ref{replicated-1},
where the states are ordered ac\-cording to the inverse binary map
$\alpha\left(\sigma_{V}\right)=1+\sum_{i\leq N}2^{N-i}\left({\textstyle \frac{1+\sigma_{i}}{2}}\right).$}
\begin{figure}
\begin{centering}
\textit{\includegraphics[scale=0.23]{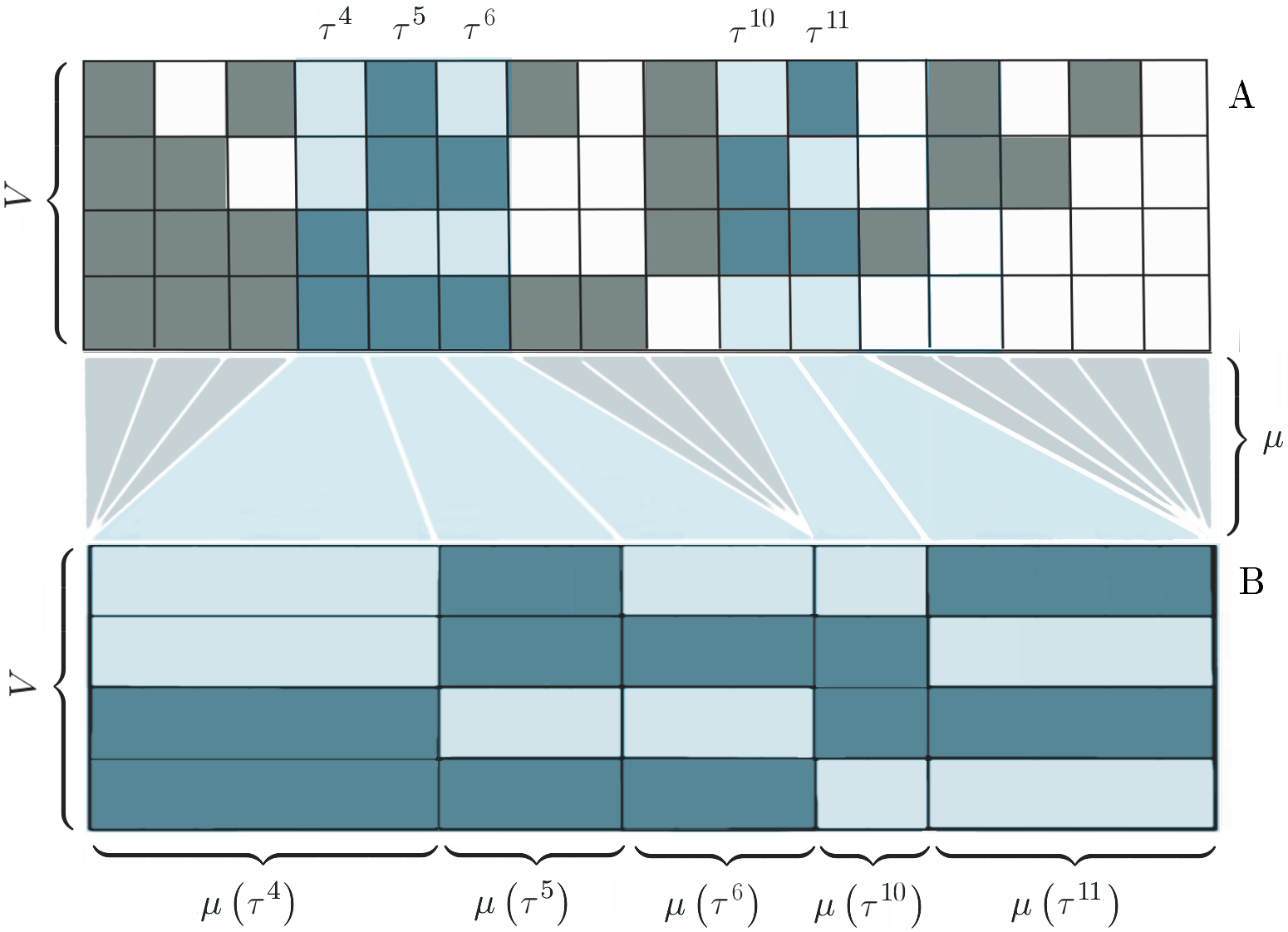}}
\par\end{centering}

\textit{~}

\centering{}\textit{\caption{\label{fig3.1}Kernel representation $M_{\mu}\left(x,y\right)$ of
Eq. (\ref{eq:magkernel}) (lower kernel B) and its support kernel
$M_{\nu}\left(x,y\right)$ (upper kernel A) for a system of $N=4$
spins described by a trial distribution with $\mu\left(\tau_{V}^{\alpha}\right)>0$
for $\alpha\in\{4,5,6,10,11\}$ and zero otherwise. Dark cells indicate
spin down, bright cells spin up. Between the two kernels it is shown
the action of $\mu$ on the support kernel $M_{\nu}$ to get the actual
kernel $M_{\mu}$. The states are ordered following the $\alpha-$index
of Definition \ref{1(Magnetization-Kernel)-Let}, then $\tau_{V}^{4}=\left(+,+,-,-\right)$,
$\tau_{V}^{5}=\left(-,-,+,-\right)$, $\tau_{V}^{6}=\left(+,-,+,-\right)$,
$\tau_{V}^{10}=\left(+,-,-,+\right)$, $\tau_{V}^{11}=\left(-,+,-,+\right)$. }
}
\end{figure}
 
\end{defn}
Array encodings of the order parameters have been considered since
the beginning in the context of the Spin Glasses theory (see the overlap
matrix of \cite{Parisi-2}), but their use to represent probability
distributions is recent enough. Before \cite{ACO-1,ACO2}, for example,
the Aldous-Hoover theorem has been used in \cite{Panchenkohoover}
to encode the replicated distribution of the SK model into a four
dimensional spin tensor. We remark that $\mu\left(\sigma_{V}\right)$
is defined up to an arbitrary reshuffling of $\alpha$, if we apply
the discrete invertible map $\alpha\rightarrow\theta\left(\alpha\right)$
still 
\begin{equation}
\sum_{\alpha\in S}\mu\,(\tau_{V}^{\theta\left(\alpha\right)})\prod_{i\in V}{\textstyle \left(\frac{1+\tau_{i}^{\theta\left(\alpha\right)}\,\sigma_{i}}{2}\right)}=\sum_{\alpha\in S}\mu\left(\tau_{V}^{\alpha}\right)\prod_{i\in V}{\textstyle \left(\frac{1+\tau_{i}^{\alpha}\,\sigma_{i}}{2}\right)}
\end{equation}
because for probability measures the labeling of the support is a
free parameter. \textit{}
\begin{figure}
\centering{}\textit{\includegraphics[scale=0.22]{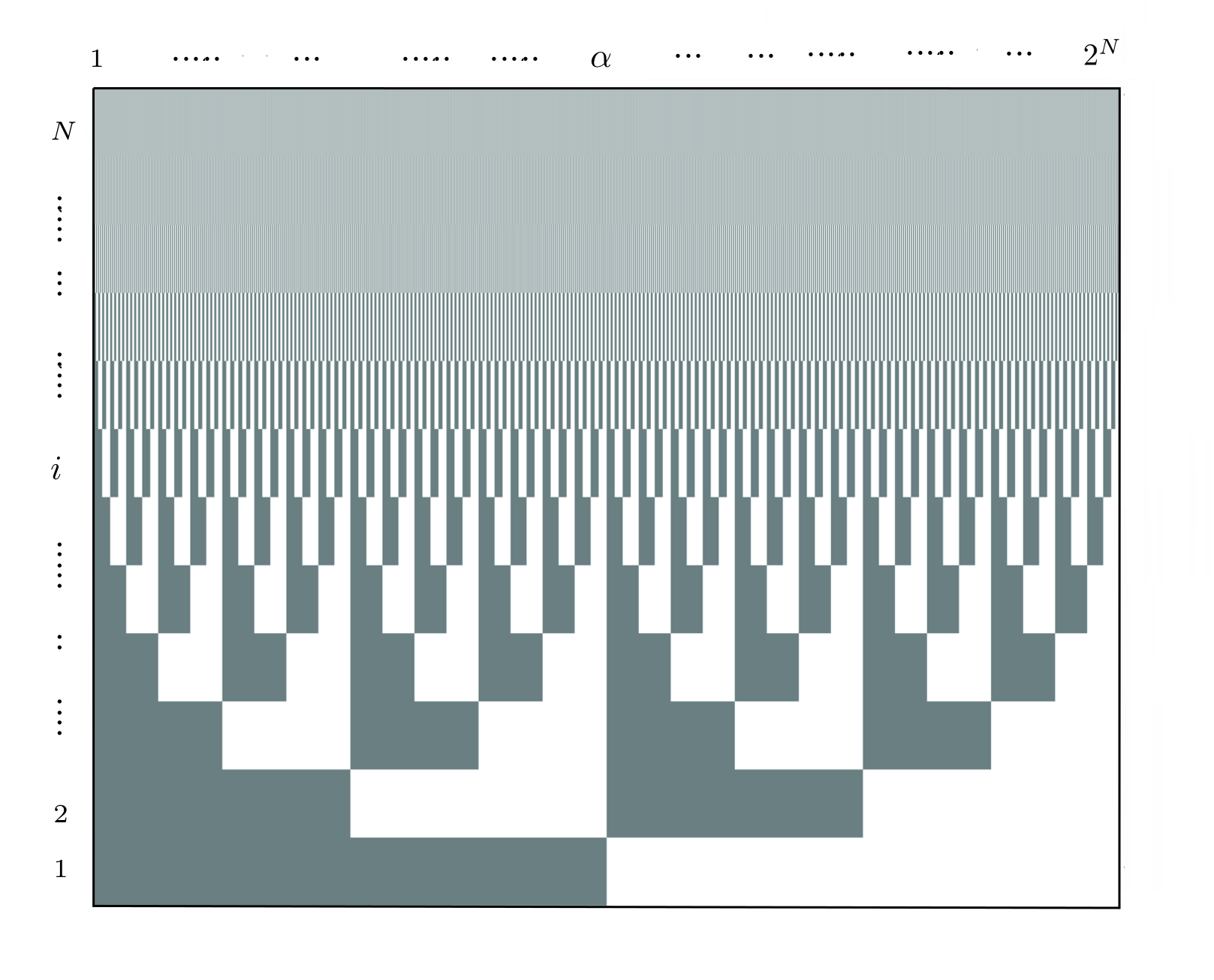}}\\
\textit{~\caption{\textit{\label{replicated-1}Support kernel $M_{\nu}\left(x,y\right)$
associated to the support (uniform) measure $\nu$ for a system of
$N=12$ spins. As before, the spin up is in bright color and the spin
down is in darker shade. The states have been disposed according to
the $\alpha-$index of Definition \ref{1(Magnetization-Kernel)-Let}
in increasing order from $\alpha=1$, that is $(-,-,\,...\,,-)$,
to $\alpha=2^{N}$, that is $(+,+,\,...\,+)$. As one can appreciate
from the figure, the index highlights a hierarchical structure that
exist between the magnetization states.}}
}
\end{figure}
\textit{}
\begin{figure}
\centering{}\textit{\includegraphics[scale=0.14]{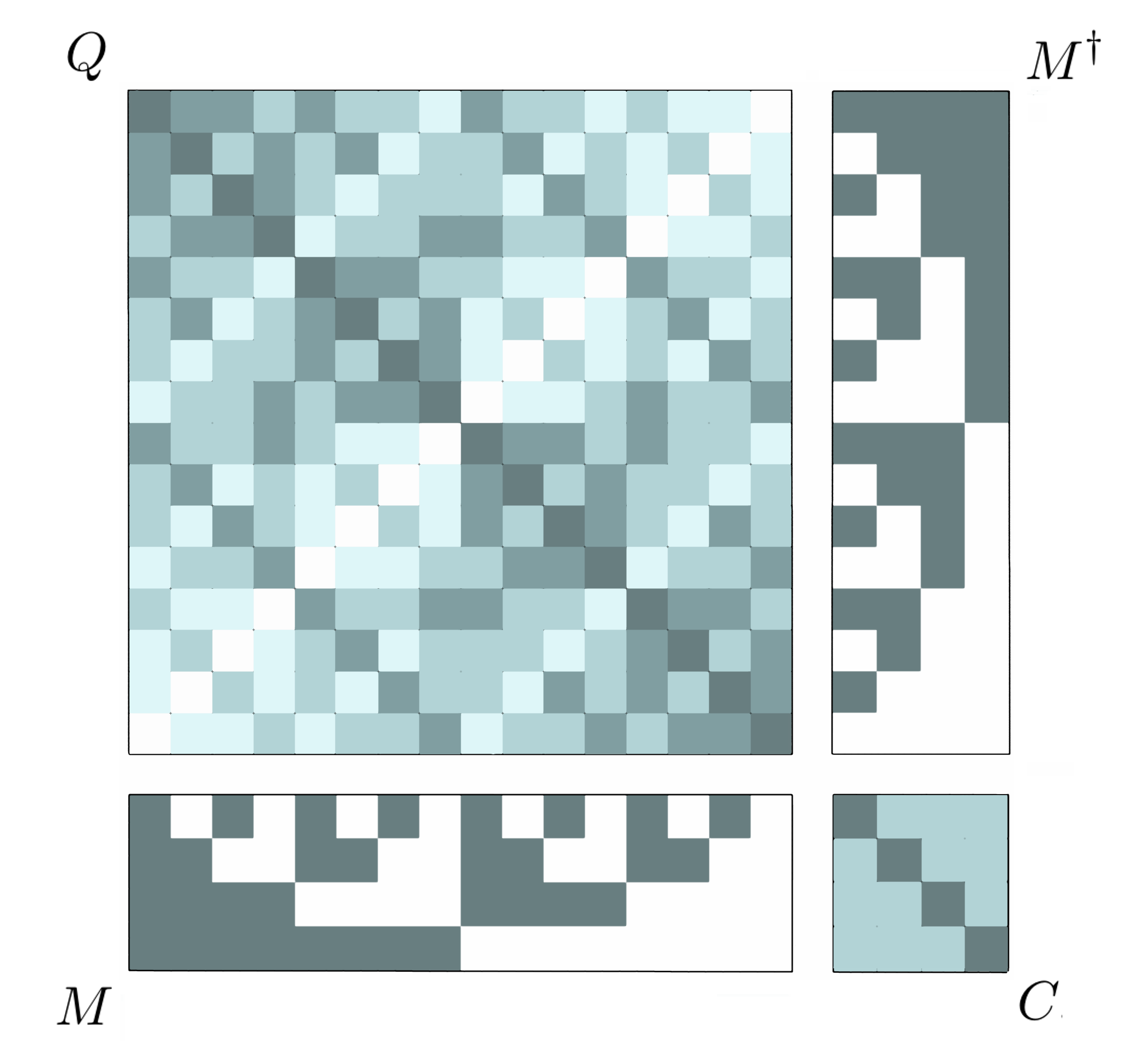}}\\
\textit{~\caption{\textit{\label{transposedkernel}} \textit{Non compactified kernel
of the support $\nu$ for a system of $N=4$ spins, and its transposed
version $\nu^{\dagger}$. The figure shows at bottom-left: $M$, a
non compactified version of the kernel of $\nu$, like in Definition
\ref{1(Magnetization-Kernel)-Let} but with $x_{i+1}-x_{i}=y_{\alpha+1}-y_{\alpha}=1$.
At top-right: $M^{\dagger}$, non-compactified kernel of the transposed
support $\nu^{\dagger}$ of Eqs. (\ref{eq:sylvester}) and (\ref{eq:transup}).
At top-left and bottom-right: $Q$ and $C$, the overlap matrix and
the correlation matrix: color shades correspond to the possible values
$1$ (darkest shade), $1/2$, $0$, $-1/2$, $-1$ (white).}}
}
\end{figure}

Then, the kernel of $\mu$ is not unique, because there are a large
number of possible choices for the map $\theta$ that points to the
same $\mu$. But, in general, the order in which the states are observed
gives informations about the pattern that minimizes the action. The
physical meaning of the index $\alpha$ (and the $\theta$ maps in
general) is best understood if the same experiment before is figured
for spin systems that are not random, for example some Lagrangian
system. In this setting two measurements are independent if taken
at time intervals that are many times larger than the recurrence time
of the system. By choosing a specific order for the states we are
then fixing a time gauge, eventually adding some momentum to the Hamiltonian
that describes the system.

Notice that the kernel function provides a powerful encoding of correlations
and overlaps (and the event algebra in general). The following lemma
express this important feature of the kernel representation. Clearly,
we can write higher order correlation fun\-ctions and overlaps using
the same procedure: 
\begin{defn}
\textit{\label{lem:(Correlations-and-Overlaps)}(Correlations and
Overlaps) Let $i,j\in V$ and select two rows of $M_{\mu}$ such that
$z_{i}\in\left(x_{i-1},x_{i}\right]$, $z_{j}\in\left(x_{j-1},x_{j}\right]$,
then the scalar product between the two rows $z_{i}$ and $z_{j}$
is the two point correlation function
\begin{equation}
\langle\,\boldsymbol{\sigma}_{i}\boldsymbol{\sigma}_{j}\rangle_{\mu}=\sum_{\tau_{V}\in S}\mu\left(\tau_{V}\right)\tau_{i}\tau_{j}=\int_{y\in\left[0,1\right]}dy\,M_{z_{i}}^{y}M_{z_{j}}^{y}.
\end{equation}
Moreover, let $\sigma_{V}$ and $\tau_{V}$ be two magnetization states,
let $t_{\sigma}\in S\left(\sigma_{V}\right)$ and $t_{\tau}\in S\left(\tau_{V}\right)$,
then the scalar product between the columns $t_{\sigma}$ and $t_{\tau}$
of the kernel $M$ is the magnetization overlap between these states
\begin{equation}
q\,(\sigma_{V},\tau_{V})=\frac{1}{N}\sum_{i\in V}\sigma_{i}\tau_{i}=\int_{x\in\left[0,1\right]}dx\,M_{x}^{t_{\sigma}}M_{x}^{t_{\tau}}.
\end{equation}
The property can be trivially verified by substituting the definition
of $M$ into the above formulas. }
\end{defn}
Notice that this last statements admit an interesting operatorial
description: if $M^{\dagger}$ is the transposed kernel, then $M^{\dagger}M=Q$
and $M_{\,}M^{\dagger}=C$, where $Q$ and $C$ are the overlap and
correlation matrices rescaled to the unitary square. In fact, according
to the Definition \ref{lem:(Correlations-and-Overlaps)} we can use
kernels to represent creation and annihilation operators by the commutator
\begin{equation}
\left[M^{\dagger},M\right]=M^{\dagger}M-M_{\,}M^{\dagger}=Q-C,
\end{equation}
where $Q$ and $C$ are the overlap and correlation matrices rescaled
to the unitary square. Two questions immediately arises, what happen
if
\begin{equation}
\left[M^{\dagger},M\right]=0
\end{equation}
in some way, and the meaning of the measure $\mu^{\dagger}$ associated
to $M^{\dagger}$. 

If the commutator is zero the overlap matrix weakly converges to the
cor\-relation matrix, then the averages on $V$ matches those on
$S$ after proper rescaling of the variable on which we take the average.
Concerning the transposed mea\-sure, it describes a whole new spin
system whose correlations and overlaps are exchanged in role respect
to $\mu$, we can give a simple definition as follows:
\begin{defn}
\textit{\label{Transposed support}(Transposed measure $\mu^{\dagger}$)
Let $\mu\in\mathcal{P}\left(\Omega^{V}\right)$ be a probability measure
de\-scri\-bing a system of $|V|=N$ spins, and let $M$ be its kernel.
Starting from $\mu$ we can de\-fi\-ne a new sequence of probability
measures $\mu_{n}^{\dagger}\in\mathcal{P}\left(\Omega^{R}\right)$,
each acting on a different spin space $\Omega^{R}$ with different
(eventually much larger) number of spins $|R|=n$, such that for $n\rightarrow\infty$
the sequence of the associated kernels weakly converges to $M^{\dagger}$.
We indicate this limit with the symbol $\mu^{\dagger}$, and call
it the transposed measure of $\mu$. }
\end{defn}
Notice that the sequence $\mu_{n}^{\dagger}$ is not unique, we can
define many that converge to the same limit kernel. To precisely describe
these concepts it would be in fact necessary to introduce the cut
distance convergence and other graph theoretic arguments that are
needed to work with limit kernels (an introduction can be found in
Section \ref{sec:Kernel-filtration}), but there are already interesting
cases in which $\mu$ is regular enough such that $\mu^{\dagger}$
can be defined also when $n$ and $N$ are finite. For example, consider
the transposed support $\nu^{\dagger}$. Let $R=S$, and consider
the spin vectors 
\begin{equation}
\sigma_{S}\in\Omega^{S},\ \ \ |\Omega^{S}|=|\Omega^{\Omega^{V}}|=2^{2^{N}},
\end{equation}
then we introduce $\mathscr{S}$, that is a collection of $N$ states
of $2^{N}$ spins
\begin{equation}
\mathscr{S}:=\left\{ \,\rho_{S}^{i}\in\Omega^{S}:\,i\in V\right\} \subset\Omega^{S},\label{eq:sylvester}
\end{equation}
these states are eventually the row vectors of $\Omega^{V}$, 
\begin{equation}
\rho_{S}^{i}:=\left\{ \rho_{\alpha}^{i}\in\Omega:\rho_{\alpha}^{i}=\tau_{i}^{\alpha},\,\alpha\in S\right\} ,
\end{equation}
and are obtained by the following construction; start from $1_{S}$,
a magnetization state with all positive spins, and flip half of them
to get two groups: one with all positive spins and one with all negative.
Then, apply this procedure iteratively inside each group, until the
state $i=N$, where the spins oscillate in sign between each $\alpha$
and $\alpha+1$ (see Figure \ref{transposedkernel}). Then, the transposed
measure $\nu^{\dagger}$ is 
\begin{equation}
\nu^{\dagger}\left(\sigma_{S}\right)=\frac{1}{N}\sum_{i\in V}\prod_{\alpha\in S}{\textstyle \left(\frac{1+\rho_{\alpha}^{i}\,\sigma_{\alpha}}{2}\right)}.\label{eq:transup}
\end{equation}
notice that due to their special construction, the states $\rho_{R}^{i}$
are exactly orthogonal, ie this set has overlap exactly zero between
any pair of states. 

What allows to define $\nu^{\dagger}$ with finite $n$ is that the
measure $\nu$ is constant, then it can be reduced to a rational number
apart from a global rescaling. For non uniform real mea\-sures we
can only write an approximating sequence: let $n>N$ and split $R$
into $2^{N}+1$ disjoint regularized subsets $R_{\alpha}$ with $\alpha\in S$,
plus one irregular $R_{0}$ that collects the real valued reminders:\textit{
\begin{equation}
|R_{\alpha}|=\left\lfloor n\,\mu\left(\tau_{V}^{\alpha}\right)\right\rfloor /n\in\mathbb{N},\ \ |R_{0}|=1-{\textstyle \sum_{\alpha}}|R_{\alpha}|.
\end{equation}
}When $n\rightarrow\infty$ the reminder becomes irrelevant, and the
sequence of measures 
\begin{equation}
\mu_{n}^{\dagger}\left(\sigma_{R}\right):=\frac{1}{N}\sum_{i\in V}\prod_{\alpha\in S}\prod_{a\in R_{\alpha}}{\textstyle \left(\frac{1+\rho_{\alpha}^{i}\,\sigma_{a}}{2}\right)}
\end{equation}
converges to $\mu^{\dagger}$. Luckily enough, the kernel we will
deal with for the SK model (ker\-nel of the eigenstates of magnetization)
can be transposed for finite $n$ like in the $\nu$ case, and we
can partially avoid the technicalities that one would need to manage
with limit kernels. 

The physical significance of these transposed kernels is in that we
in\-ter\-pret them as tho\-se that actually describe the 1-RSB
phase of the Parisi Ansatz, in fact, we interpret the Parisi full-RSB
ansatz as a way to split the systems into sub-systems who\-se kernel
weakly commute in the thermodynamic limit, so that the averages can
be done with the transposed measure. We remark that the tran\-sposed
measure is typically defined on an exponentially larger set of spins:
this suggests a connection with the replicated system. 

In the Section \ref{sec:The-Ansatz} of this paper, we will show that
for SK a possible scheme to obtain su\-ch par\-tition into commuting
sub-systems is to simply split the spin group into small sub\-groups
and apply the Bayes rule. The physical idea behind is in that any
probability measure describing an actual physical spin system defined
for variable number of spins must be coherent with the fact that such
system has been constructed or created in some way. Then, it must
always be possible to con\-struct a sequence of systems of increasing
size that eventually converge to the one we are looking at. 

Consider, for example, an Ising Model in $d=3$ in which $N$ spins
are arranged into a cube: such system can be constructed starting
with a single spin, then ad\-ding a layer of nearest neighbors, then
add another and so on, until reaching the size $N$. Notice that this
idea is not new at all, being the same that motivates the Cavity methods
and the Gr\-and Canonical ensemble. 

The most convenient layering scheme depends on the system, but in
general each layer has two kinds of energy contributions: those between
the spin of the layer itself, that we call the \textit{core }contribution,
and those between the layer and the previous ones (just the previous
in case of the finite dimensional Ising model before) that we call
the \textit{interface}. 

\begin{figure}[p]
\centering{}\textit{\includegraphics[scale=0.195]{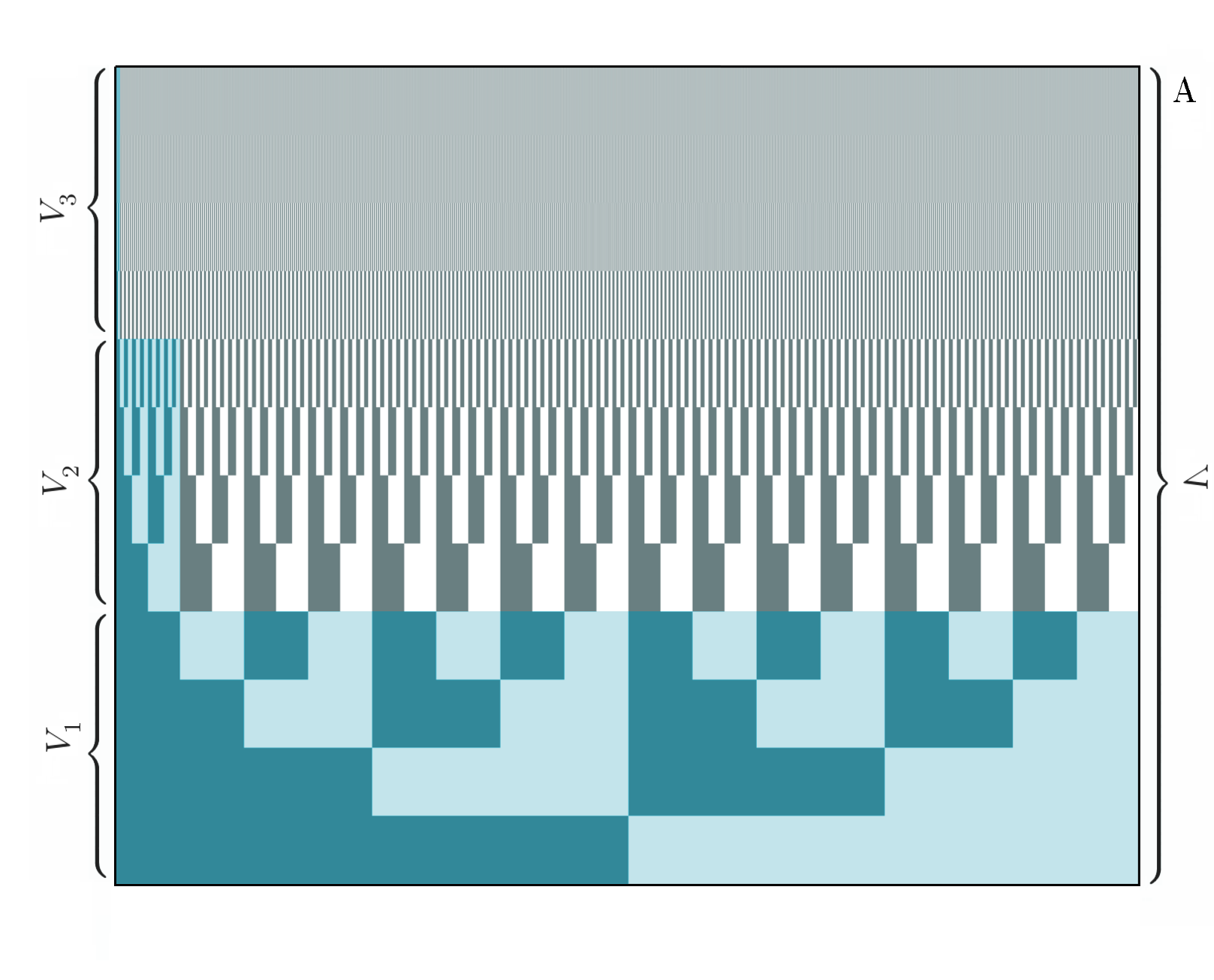}}\\
\textit{~\includegraphics[scale=0.19]{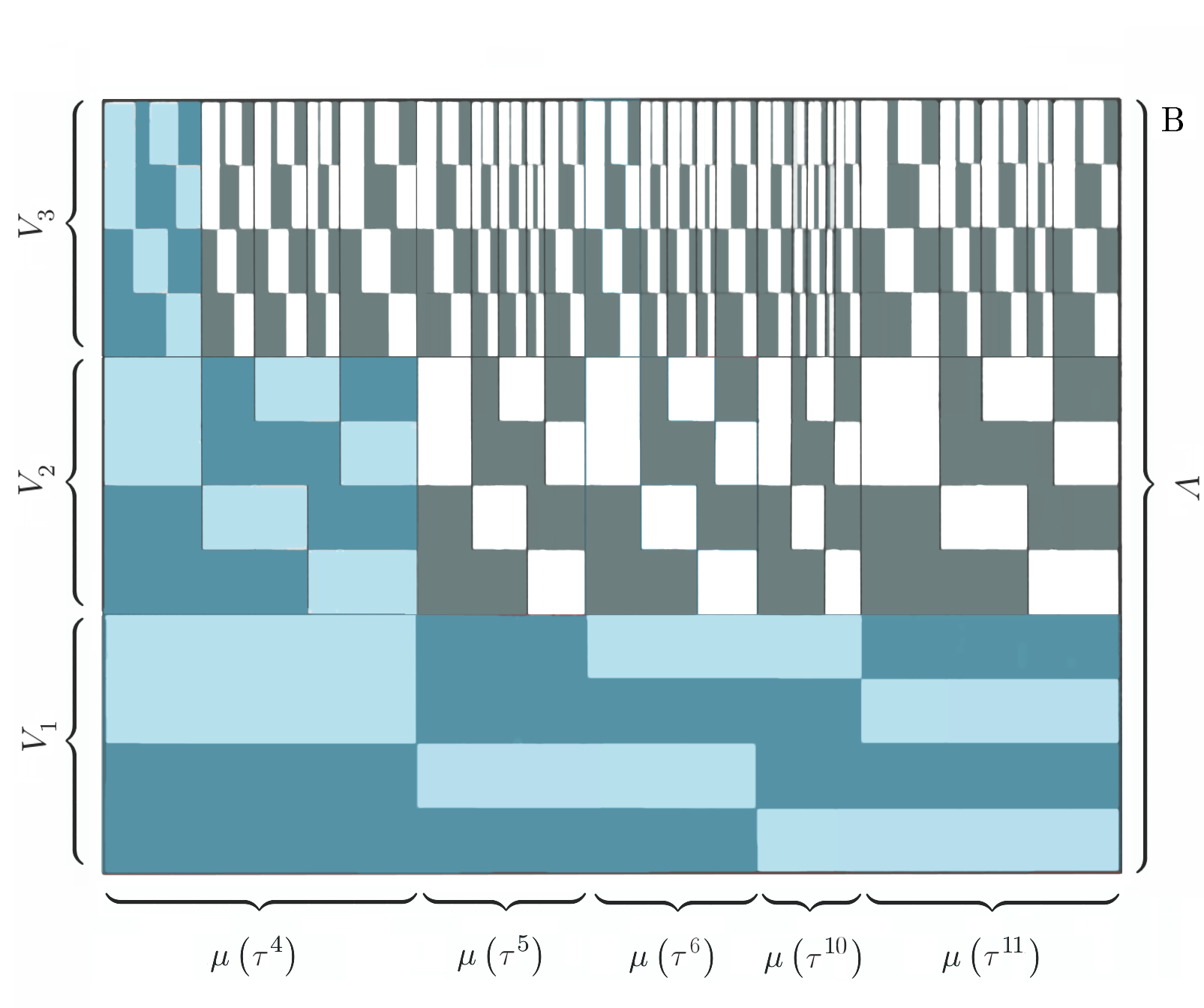}\caption{\textit{\label{replicated}Kernel of a product measure $\mu_{V_{1}}\otimes\mu_{V_{2}}\otimes\mu_{V_{3}}$
(lower kernel B) and its support (upper kernel A). Here we show the
special case of three replicas of the same measure $\mu$ of Figure
\ref{fig3.1} located at $V_{1}$, $V_{2}$ and $V_{3}$, ie we take
$\mu_{V_{1}}=\mu_{V_{2}}=\mu_{V_{3}}=\mu$ (replicated kernel).}}
}
\end{figure}

\begin{figure}[p]
\includegraphics[scale=0.155]{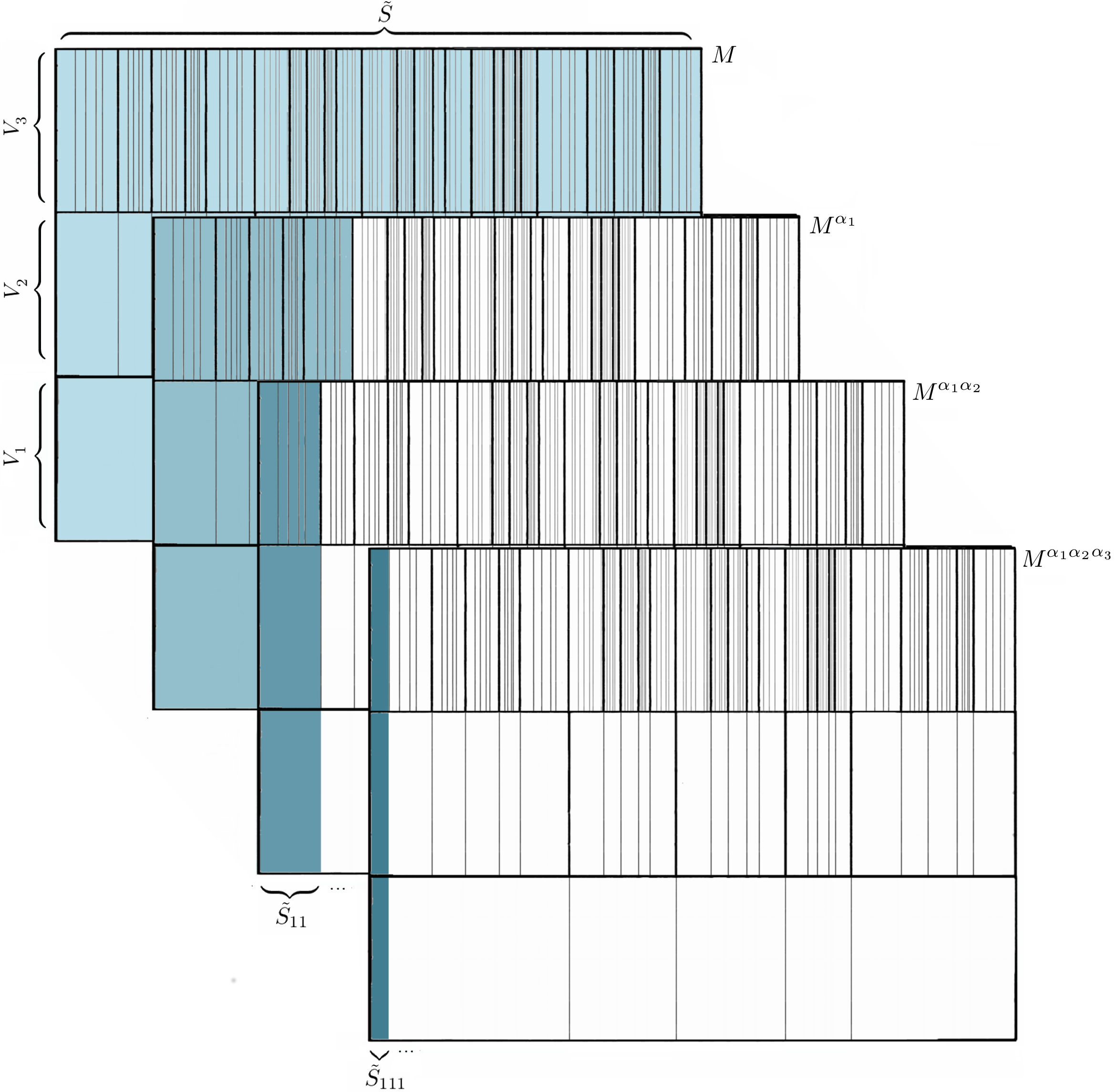}\caption{\label{fig4.1-2}Kernel representation of the filtration process according
to $\mathscr{S}\left(\mathcal{V}\right)$ of Definition \ref{11(Filtration-of-).}
for the same measure $\mu$ of Figure \ref{replicated}. The vertical
lines highlight the pure states of each layer $M$, $M^{\alpha_{1}}$
and $M^{\alpha_{1}\alpha_{2}}$ and the last kernel is $M_{\mu}$
itself, the first pure state of each level is highlighted in blue.}
\end{figure}

\begin{figure}
\includegraphics[scale=0.195]{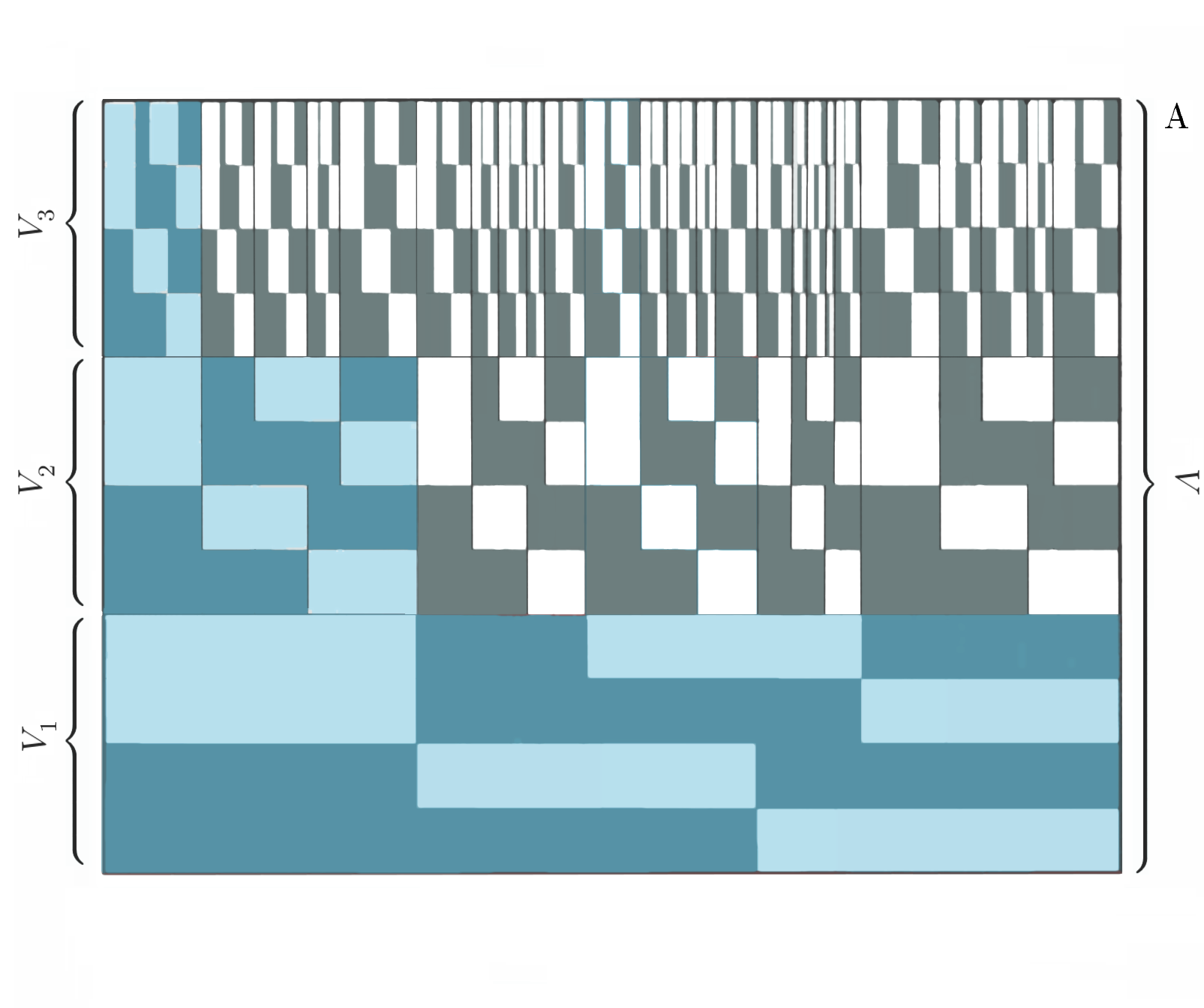}

\includegraphics[scale=0.195]{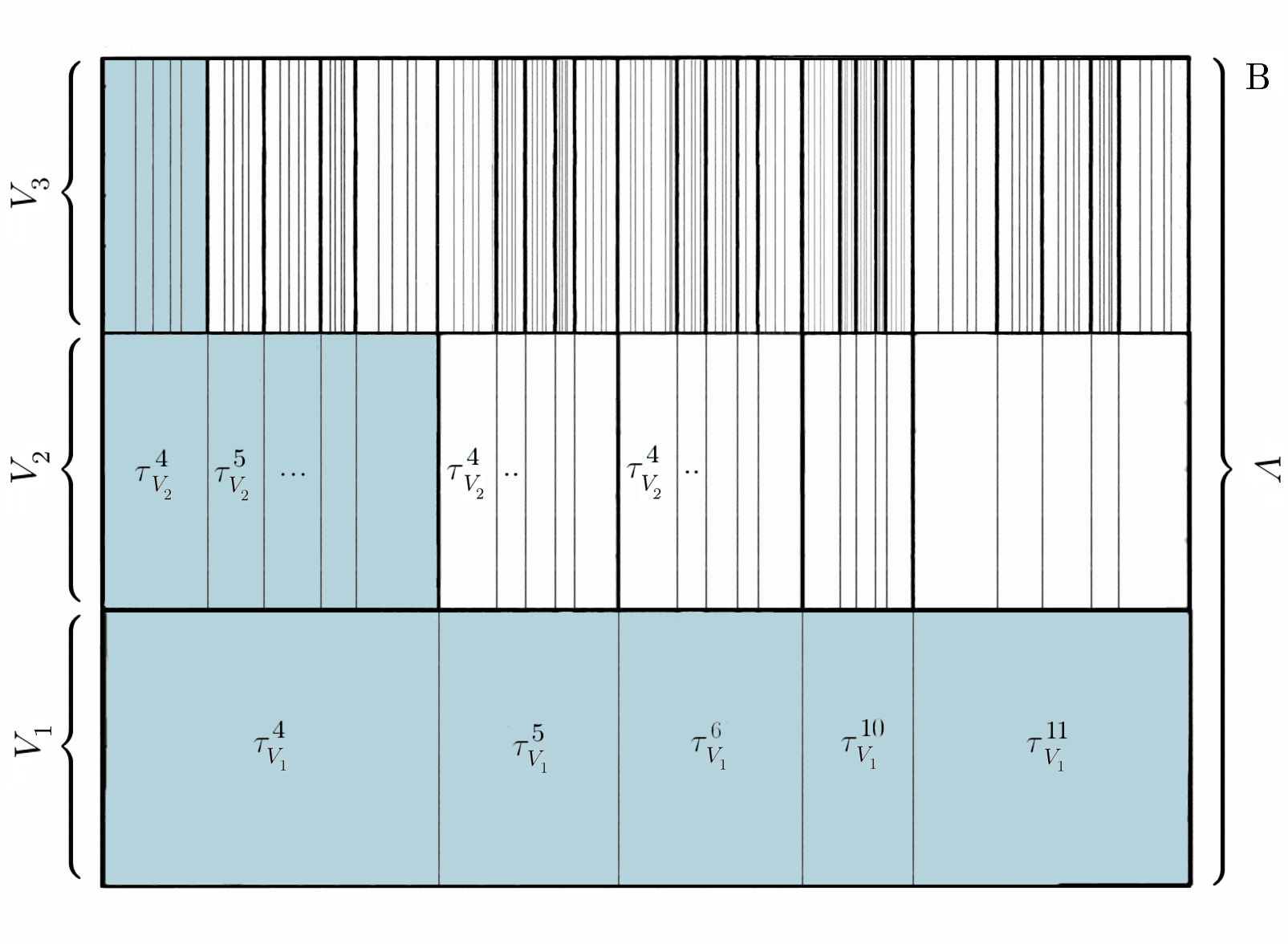}

\caption{\label{fig4.1} Binary index of Definition \ref{1(Magnetization-Kernel)-Let}
applied to the same kernel of Figure \ref{replicated} (upper kernel
A). In the lower kernel B we explicitly show the states classified
according to the $\alpha-$index. }
\end{figure}

\begin{figure}
\includegraphics[scale=0.195]{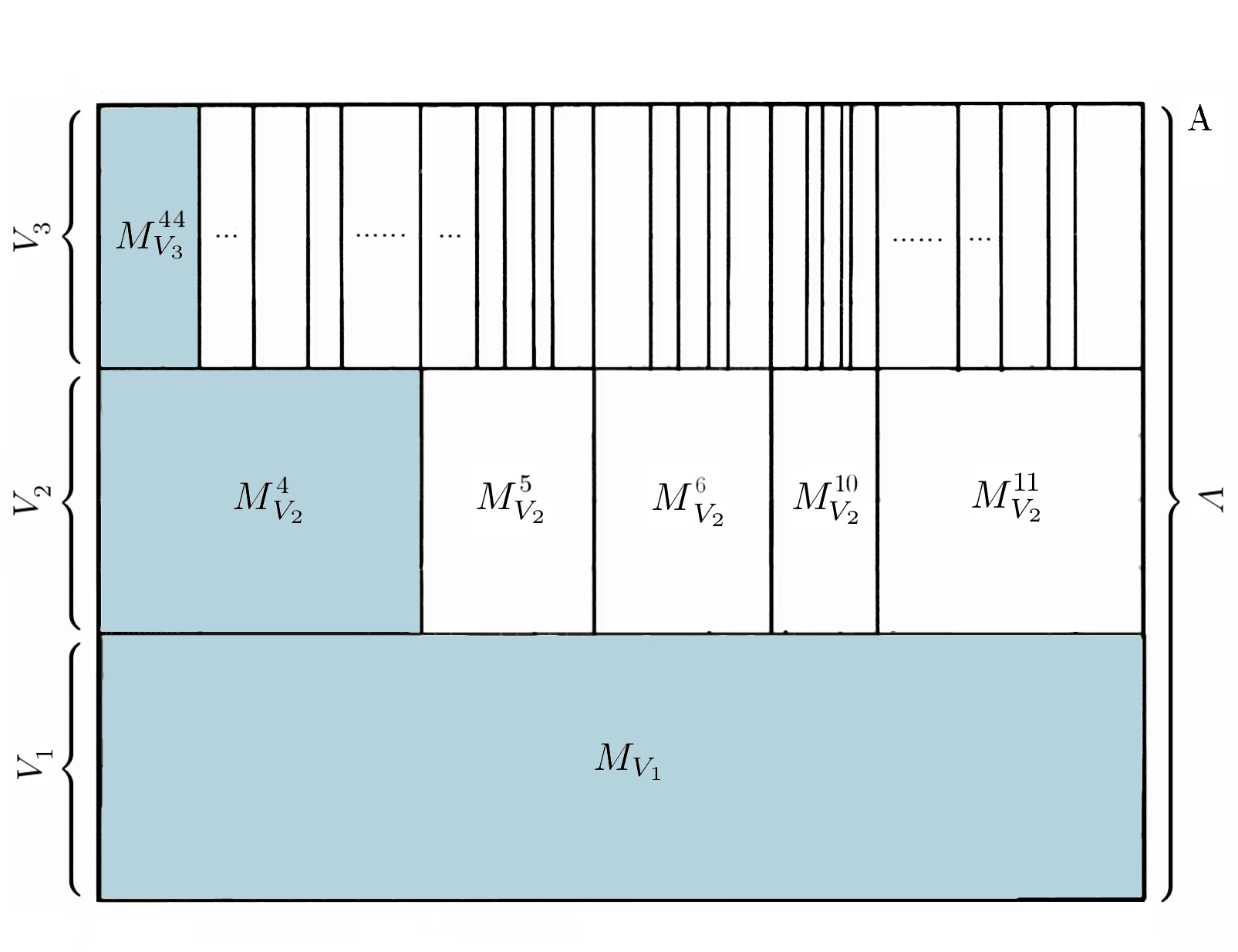}

\includegraphics[scale=0.195]{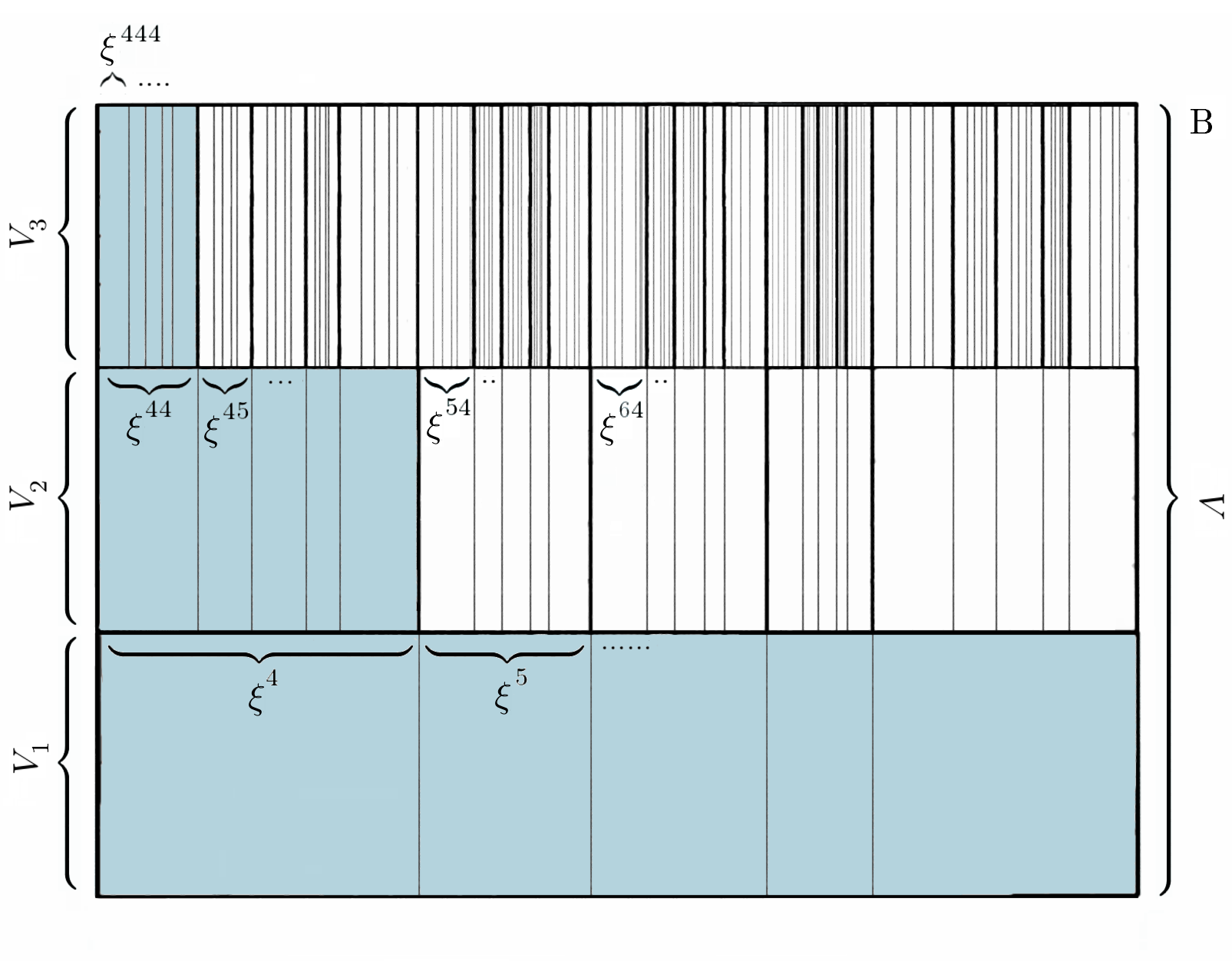}

\caption{\label{fig4.1-1}Pure state layers of Definition \ref{14(Pure-States)-Let}
in kernel representation for the same $\mu$ of Figure \ref{replicated}
and \ref{fig4.1}. The upper kernel A shows the locations of the Pure
state layers $M_{V_{\ell}}^{\alpha_{1}...\alpha_{\ell-1}}$, while
in the lower kernel B we show the refinements $S_{\alpha_{1}...\alpha_{\ell-1}}$,
highlighted by their weights $\xi^{\alpha_{1}...\alpha_{\ell-1}}$. }
\end{figure}

In the case of fully connected models, we expect that if such partition
is into very ti\-ny layers, then the contributions from the core
can be ignored respect to that of the in\-ter\-face, and that this
will make the layer commute. In fact, notice that in case of the SK
model (but the same holds for the Curie-Weiss model) there is no space
structure, and any spin that is added form a layer itself, with a
large interface. Following \cite{Franchini}, we can formally define
our analogue of the pure states of the RSB ansatz by partitioning
the vertex set $V$ into $L$ subsets $V_{\ell}$,
\begin{equation}
\mathcal{V}=\left\{ V_{1},V_{2},\,...\,,V_{L}\right\} ,
\end{equation}
we label the parts by the ordered index $1\leq\ell\leq L$ and also
relabel the vertexes inside each $V_{\ell}$, for $i\in V_{\ell}$
we apply a map $i\rightarrow i_{\ell}$ such that $1\leq i_{\ell}\leq|V_{\ell}|$.
We will refer to 
\begin{equation}
\sigma_{V_{\ell}}=\left\{ \sigma_{i}\in\Omega:\,i\in V_{\ell}\right\} 
\end{equation}
as layer magnetization states. Also, it will be convenient to express
$\mathcal{V}$ in terms of the sequence $Q_{\ell}$, with $|Q_{\ell}|=q_{\ell}|V|$
and $0\leq q_{\ell}\leq1$. Starting from $Q_{1}=V_{1}$ this sequence
is defined recursively $Q_{\ell}=\bigcup_{t\leq\ell}V_{t}$ until
the last step $\ell=L$, corresponding to the whole vertex set. The
associated sequence of states is as follows:
\begin{equation}
\sigma_{Q_{\ell}}=\bigcup_{t\leq\ell}\sigma_{V_{\ell}}\in\Omega^{Q_{\ell}},
\end{equation}
they are composed by the first $\ell$ sub-states $\sigma_{V_{\ell}}$. 
\begin{defn}
\textit{\label{11(Filtration-of-).}(Filtration of $S$ induced by
$\mathcal{V}$) Let define the subsets 
\begin{equation}
S\left(\sigma_{V}\right)=\left\{ \sigma_{V}\right\} 
\end{equation}
each composed by one element of $S$. Then, the filtration of $S$
induced by $\mathcal{V}$ 
\begin{equation}
\mathscr{S}\left(\mathcal{V}\right)=\left\{ \mathcal{S}_{\ell}\left(\mathcal{V}\right):\,1\leq\ell\leq L\right\} 
\end{equation}
is defined as the sequence of partitions 
\begin{equation}
\mathcal{S}_{\ell}\left(\mathcal{V}\right)={\textstyle \left\{ S\left(\sigma_{Q_{\ell}}\right)\subseteq S:\,\sigma_{Q_{\ell}}\in\Omega^{Q_{\ell}}\right\} },
\end{equation}
that is obtained by recursively joining the subsets according to the
iteration 
\begin{equation}
S\left(\sigma_{Q_{\ell-1}}\right)=\bigcup_{\sigma_{V_{\ell}}\in\Omega^{V_{\ell}}}\,S\left(\sigma_{Q_{\ell}}\right),
\end{equation}
down to $\ell=1$. Let $\mu\in\mathcal{P}\left(\Omega^{N}\right)$
and take some partition $\mathcal{V}$, we write
\begin{equation}
\mu_{\ell}\left(\sigma_{Q_{\ell}}\right)=\sum_{\tau_{V}\in S(\sigma_{Q_{\ell}})}\mu\left(\tau_{V}\right)
\end{equation}
for the probability mass of $S\left(\sigma_{Q_{\ell}}\right)$ under
$\mu$}.\textit{ Let $f:\,\Omega^{V}\rightarrow\mathbb{R}$, then,
the average va\-lue $\langle f\left(\boldsymbol{\sigma}_{V}\right)\rangle_{\mu}$
according to $\mu$ is obtained starting from
\begin{equation}
f_{L}\left(\sigma_{Q_{L}}\right)=f\left(\sigma_{Q_{L}}\right),
\end{equation}
where $Q_{L}=V$, then we iterate the formula backward 
\begin{equation}
f_{\ell-1}\left(\sigma_{Q_{\ell-1}}\right)=\sum_{\sigma_{V_{\ell}}\in\Omega^{V_{\ell}}}\xi_{\ell}\left(\sigma_{Q_{\ell}}\right)\,f_{\ell}\left(\sigma_{Q_{\ell}}\right),\label{eq:recursin}
\end{equation}
the average is taken according to the following distribution
\begin{equation}
\xi_{\ell}\left(\sigma_{Q_{\ell}}\right):=\mu_{\ell}\left(\sigma_{Q_{\ell}}\right)/\mu_{\ell-1}\left(\sigma_{Q_{\ell-1}}\right),
\end{equation}
that is the distribution of the layer $\sigma_{V_{\ell}}$ for a given
$\sigma_{Q_{\ell-1}}$ }\cite{Franchini}\textit{.}
\end{defn}
\noindent This is enough to analyze the SK Hamiltonian, but before
that, it will be useful to dis\-cuss the kernel $M$, shown in Figures
\ref{replicated}, \ref{fig4.1-2}, \ref{fig4.1}, \ref{fig4.1-1}.
We identify our analogue of the RSB pure states in the sub-kernels
associated to the partitions $\mathcal{S}_{\ell}\left(\mathcal{V}\right)$
(see below). 

We anticipate that the following definition aims to generalize the
concept of pu\-re state to any spin distribution, and does not yet
have all the properties of the construction that one finds in \cite{Talagrand},
which we refer to as SK pure states and discuss in Section \ref{sec:The-Ansatz}.
\begin{defn}
\textit{\label{14(Pure-States)-Let-1}(Analogue Pure States) Let $\mathscr{S}$$\left(\mathcal{V}\right)$
be a filtration induced by the partition $\mathcal{V}$ as in Definition
\ref{11(Filtration-of-).}, let $\mathcal{S}_{\ell}\left(\mathcal{V}\right)$
be the partition associated to the $\ell-$th level of refinement,
and let $M$ the kernel associated to $\mu$. Then, we can identify
a partition 
\begin{equation}
{\textstyle M_{x}^{y}(\sigma_{Q_{\ell-1}})=M_{x}^{y}\cdot I[\,y\in\tilde{S}\left(\sigma_{Q_{\ell-1}}\right)]}.
\end{equation}
where $\tilde{S}\left(\sigma_{Q_{\ell-1}}\right)$ is the image of
$S\left(\sigma_{Q_{\ell-1}}\right)$ on $[0,1]$. Hereafter will refer
to these sub-kernels as the Pure States of $M$ according to $\mathcal{S}_{\ell}\left(\mathcal{V}\right)$. }
\end{defn}
The analogue pure states of the $\ell$-th level are identified with
the partition that one gets after $\ell$ refinements of $S$, an
example is in Figure \ref{fig4.1-2}, the sub-kernel associated to
the first pure state of each level is highlighted in blue.\textit{}

Notice that for any pair of $\sigma_{V},\tau_{V}\in S\left(\sigma_{Q_{\ell}}\right)$,
holds that the overlap (scalar product) between the magnetization
states $\sigma_{V}$ and $\tau_{V}$ satisfy the inequality $\sigma_{V}\cdot\tau_{V}\geq|Q_{\ell}|-|V\setminus Q_{\ell}|$
because by definition $\sigma_{i}=\tau_{i}$ at least for any $i\in Q_{\ell}$.
By \cite{Talagrand}, any overlap dis\-tri\-bution inside an SK
pure state is expected to con\-cen\-tra\-te on some nontrivial
value for large systems, this is recovered under the ad\-di\-tio\-nal
assumption that $\sigma_{i}$ and $\tau_{i}$ behaves independently
for $i\in V\setminus Q_{\ell}$, gi\-ving $\sigma_{V}\cdot\tau_{V}=|Q_{\ell}|+o(N)$
almost surely. 

Although the previous definition allows to connect with the usual
objects of Spin Glass theory, this partition structure of $M$ is
not the most natural that one can arrange. In the following we define
a second version of the pure states, which we call Layer States. These
are not directly related with the usual notion of Pure State that
is found in Spin Glass (SG) literature, and we interpret them as the
transposed version the 1RSB pure states.
\begin{defn}
\textit{\label{14(Pure-States)-Let}(Layer states) Let $\mathscr{S}$$\left(\mathcal{V}\right)$
be a filtration induced by the partition $\mathcal{V}$ as in Definition
\ref{11(Filtration-of-).} and let $M$ the kernel associated to $\mu$.
We can identify a partition of $M$ into the sub-kernels associated
to the $\xi_{\ell}\left(\sigma_{Q_{\ell}}\right)$ distributions 
\begin{equation}
{\textstyle M_{x}^{y}\left(\sigma_{Q_{\ell-1}}\right)=M_{x}^{y}\cdot I[\,x\in\tilde{V}_{\ell},\,y\in\tilde{S}\left(\sigma_{Q_{\ell-1}}\right)]}.
\end{equation}
where $\tilde{V}_{\ell}$ and $\tilde{S}\left(\sigma_{Q_{\ell-1}}\right)$
are the images of $V_{\ell}$ and $S\left(\sigma_{Q_{\ell-1}}\right)$
on the interval $[0,1]$, we will re\-fer to these sub-kernels as
the Layer States of $M$ according to $\mathscr{S}\left(\mathcal{V}\right)$.}
\end{defn}
\noindent One can confront the kernel of Figure \ref{fig4.1}A with
its partition according to the previous definition, Figure \ref{fig4.1-1}. 

The next section contains an introduction to more advanced kernel
methods, the reader mostly interested in the physics of the SK model
can jump directly to Section \ref{sec:The-Ansatz} for the moment.


\pagebreak{}

\section{Kernel filtration\label{sec:Kernel-filtration}}

We continue by presenting an alternative approximation scheme that
is intended to give a formulation for the \textit{finite volume pure
states} of Marinari et al. in \cite{Marinari} that is also compatible
with the kernel methods presented in \cite{ACO-1,ACO2,ACO3}, which
allows to operate directly in the thermodynamic limit (TL). In principle,
this representation is more general, as it is not based on any special
filtration, and in fact contains the one we used for the computations
of the SK model as special case. The connection between kernels and
the pure states of the RSB ansatz has been first noticed in \cite{ACO-1},
where a kernel encoding of $\mu$ is introduced in order to prove
the following 
\begin{lem*}
\label{(Bapst,-Coja-Oghlan,-2016)-1}(Bapst, Coja-Oghlan, 2016) For
any measure $\mu\in\mathcal{P}\left(\Omega^{V}\right)$ it is possible
to take some arbitrary small $\epsilon>0$ and a partition of $\Omega^{V}$
into a finite number $n\geq n\left(\epsilon,|K|\right)$, not dependent
from $N$, of disjoint subsets $S_{\alpha}$, $0\leq\alpha\leq n$
such that $\mu\left(S_{0}\right)\leq\epsilon$ and 
\begin{equation}
\sum_{K\in\left\{ 1,\,...\,N\right\} ^{|K|}}\left\Vert \mu_{K}^{\alpha}-{\textstyle \bigotimes_{i\in K}}\,\mu_{i}^{\alpha}\right\Vert _{TV}\leq\epsilon\,N^{\,|K|},\ \forall\,\alpha,|K|\geq1\label{eq:DECOMP1-2}
\end{equation}
if $N$ is chosen large enough (we denoted by $\left\Vert \,\cdot\,\right\Vert _{TV}$
the total variation\footnote{Given two measures $\mu\,,\nu:S\rightarrow\left[0,1\right]$ and some
$A\subseteq S$ the \textit{total variation distance} between $\mu$
and $\nu$ is given by the formula $\left\Vert {\textstyle \mu-\nu}\right\Vert _{TV}=2\,\sup_{A}|\mu(A)-\nu(A)|$.}). For example, in the case $|K|=2$ we can write 
\begin{equation}
\sum_{\left\{ i,j\right\} \in\left\{ 1,\,...\,N\right\} ^{2}}\left\Vert {\textstyle \mu_{\left\{ i,j\right\} }^{\alpha}-\mu_{i}^{\alpha}}\otimes\mu_{j}^{\alpha}\right\Vert _{TV}\leq\epsilon\,N^{2},\ \forall\,\alpha.\label{eq:DECOMP1-1-1}
\end{equation}
\end{lem*}
\begin{proof}
It is essentially a measure theoretic version of the Szemeredi Regularity
Lemma, see Chapter 9.2 and 9.3 of \cite{Lovasz} for a detailed review.
A proof of Eq.s (\ref{eq:DECOMP1-2}) and (\ref{eq:DECOMP1-1-1})
can be found in the first part of \cite{ACO-1}, after the statements
of Theorem 2.2 and Corollaries 2.3-2.5.
\end{proof}
This result tells us that for any measure $\mu$ that describes a
system of variables with finite set $\Omega$ of inner states we can
decompose our sample space $\Omega^{V}$ into a finite number $n\left(\epsilon,|K|\right)$
of regular disjoint subsets $S_{\alpha}$, $1\leq\alpha\leq n\left(\epsilon,|K|\right)$
plus one irregular $S_{0}$ with $\mu\left(S_{0}\right)\leq\epsilon$
such that for any regular subset $S_{\alpha}$ the layers of $\mu^{\alpha}$
over a randomly chosen set $K$ can be approximated by a product measure
in the sense of Eq. (\ref{eq:DECOMP1-2}). Surprisingly, the number
$n\left(\epsilon,|K|\right)$ of such regular subsets only depends
on $|K|$, $\left|\Omega\right|$ and the level of precision $\epsilon$
we want to achieve for our approximation, and it does not depend on
the size $N$ of the system. This and many other results can be obtained
by noticing that both probability measures and graphs can be exactly
encoded into kernel functions. For example, in \cite{ACO-1,ACO2,ACO4}
a new distance on $\mathcal{P}\left(\Omega^{V}\right)$ based on Graph
Theory is introduced to characterize Gibbs Measures directly in the
thermodynamic limit.
\begin{defn*}
\textit{(Cut Distance) Let $M$, $W$ be two kernels and let $\theta=(\theta_{1},\theta_{2})$
be a pair of measure preserving maps. We call Cut Norm the positive
quantity
\begin{equation}
\left\Vert M\right\Vert _{\square}=\sup_{A,B\subseteq\left[0,1\right]}\left|\int_{x\in A}\int_{y\in B}dxdy\,M\left(x,y\right)\right|
\end{equation}
and define the Cut Distance as 
\begin{equation}
D_{\square}\left(M,W\right)=\inf_{\theta}\left\Vert M-W^{\theta}\right\Vert _{\square},
\end{equation}
where $W^{\theta}$ stands for $W(\theta_{1}(x),\theta_{2}(y))$ \cite{Lovasz}.
In the context of probability theory the cut distance between $\mu\,,\nu:S\rightarrow\left[0,1\right]$
is the cut distance $D_{\square}(M_{\mu},M_{\nu})$ be\-twe\-en
the associated kernels $M_{\mu}$, $M_{\nu}$ of Eq.(\ref{eq:magkernel})
below. }
\end{defn*}
It can be shown \cite{Lovasz} that the kernel space is compact in
the Cut distance, and that convergence in cut distance is stronger
than the weak{*} convergence when dea\-ling with intensive quantities,
such as the free energy density associated to a Gibbs measure (see
Chapter 8 of \cite{Lovasz} and therein, or the first part of \cite{ACO-1,ACO2},
see Chapter 8.2 of \cite{Lovasz}, or \cite{ACO2} for the measure
theoretic approach).

The above Lemma is in fact a probabilistic version of the Szemeredi
Regularity Lemma (Chapter 9.2 and 9.3 of \cite{Lovasz}). Since the
arguments presented in the following do not require the use Szemeredi
Partitions we won't discuss this here, but we stress that these are
useful mathematical concepts and we warmly advice the reader to look
at \cite{DiaconisJansen,Lovasz} for further reading on this important
subject. 

We can give an intuitive picture by considering two independent and
equitable par\-titions of $S$ and $V$ into sub-sets $S_{a}$ and
$V_{\ell}$, their number being $n$ and $L$ respectively. Then,
define the magnetization averages inside the blocks
\begin{equation}
m_{\ell}^{a}=\frac{1}{|V_{\ell}|}\sum_{i\in V_{\ell}}\langle\boldsymbol{\sigma}_{i}\rangle_{\mu^{a}}=\frac{1}{|V_{\ell}||S_{a}|}\sum_{i\in V_{\ell}}\sum_{\alpha\in S_{a}}\mu^{\alpha}\tau_{i}^{\alpha}
\end{equation}
Then, let $\eta_{\ell}^{a}:\Omega^{V_{\ell}}\rightarrow\left[0,1\right]$
and $\eta:\Omega^{V}\rightarrow\left[0,1\right]$ be defined as follows
\begin{equation}
\eta_{\ell}^{a}\left(\sigma_{V_{\ell}}\right):=\prod_{i\in V_{\ell}}\left(\frac{1+m_{\ell}^{a}\sigma_{i}}{2}\right),\ \ \eta\left(\sigma_{V}\right):=\frac{1}{n}\sum_{a\leq n}\prod_{\ell\leq L}\eta_{\ell}^{a}\left(\sigma_{V_{\ell}}\right).
\end{equation}
Szemeredi lemma guarantees that, for any small $\epsilon$, if $n$
and $L$ are taken large enough it is possible to find a kernel such
that $D_{\square}(M_{\mu},M_{\eta})\leq\epsilon$. This fact can already
have some interesting applications because the number of parameters
that controls the trial measure $\eta$ is finite, although could
be very large if we require $\epsilon$ to be very small. For example,
let $H\left(\sigma_{V}\right)$ be some Hamiltonian and $Z$ the partition
function. By definition
\begin{multline}
Z:=\sum_{\sigma_{V}\in\Omega^{V}}e^{-\beta H\left(\sigma_{V}\right)}=\langle\,\exp\left[-\beta H\left(\sigma_{V}\right)-\log\mu\left(\sigma_{V}\right)\right]\rangle_{\mu}\leq\\
\leq\exp\left[-\beta\langle H\left(\sigma_{V}\right)\rangle_{\mu}-\langle\log\mu\left(\sigma_{V}\right)\rangle_{\mu}\right]=\exp\left[-\beta\mathcal{F}_{\beta,H}\left(\mu\right)\right],\label{eq:ssdffff}
\end{multline}
where in the second row we applied Jensen inequality, and also introduced
the Gib\-bs free energy functional
\begin{equation}
\mathcal{F}_{\beta,H}\left(\mu\right):=\langle H\left(\sigma_{V}\right)\rangle_{\mu}+\beta^{-1}\langle\log\mu\left(\sigma_{V}\right)\rangle_{\mu}.
\end{equation}
We easily obtained a variational bound for the free energy that is
optimized by the Gibbs measure $\mu^{*}$, that is 
\begin{equation}
F=-\beta^{-1}\log Z=\inf_{\mu\in\mathcal{P}\left(\Omega^{V}\right)}\mathcal{F}_{\beta,H}\left(\mu\right).
\end{equation}
Nonetheless, the function $\mathcal{F}_{\beta,H}\left(\mu\right)$
may be hard to handle, because the number of pa\-ra\-me\-ters that
controls $\mu$ grows exponentially with the size of the system, it
is much simpler is to minimize $\mathcal{F}_{\beta,H}\left(\eta\right)$
on the $nL$ pa\-ra\-me\-ters that control $\eta$,
\begin{equation}
F'=\inf_{M}\,\mathcal{F}_{\beta,H}\left(\eta\right).
\end{equation}
It is possible to show that if the cut distance $D_{\square}(M_{\mu},M_{\eta})\rightarrow0$,
then also the free energy den\-si\-ties $F'/N\rightarrow F/N$ are
weakly convergent.
\begin{defn}
\textit{(Tree Index for S and V) Let $0\leq t\leq T$, then, let introduce
the following pair of tree indexes. The first is 
\begin{equation}
A_{t}:=\alpha_{1}\alpha_{2}...\alpha_{t}\in{\textstyle \prod_{\,k\leq t}}\left\{ 1,2,\,...\,,n_{k}\right\} ,
\end{equation}
where each subindex $\alpha_{t}$ runs from $1$ to some integer $n_{t}$.
The second is 
\begin{equation}
I_{t}:=i_{1}i_{2}...i_{t}\in{\textstyle \prod_{\,k\leq t}}\left\{ 1,2,\,...\,,v_{k}\right\} .
\end{equation}
The integer parameters $n_{t}$ and $v_{t}$ are assumed the same
for each level $t$. Let de\-fine the sets $S_{A_{T}}=\left\{ A_{T}\left(\alpha\right)\right\} $
each composed by only one element of $S$ mapped onto the Tree Index
$A_{T}$ by the invertible map $A_{T}\left(\alpha\right)$. Then we
call $\mathscr{S}$, filtration of $S$, the sequence of equitable
refinements
\begin{equation}
\mathcal{S}_{t}=\left\{ S_{A_{t}}\subset S:\,A_{t}\in{\textstyle \prod_{\,k\leq t\,}}\{1,2,\,...\,,n_{k}\}\right\} ,
\end{equation}
obtained from joining the subsets $S_{A_{t}}\subseteq S$ from the
last layer $S_{A_{T}}$ down to the root level $t=0$, associated
to $S$. Then, define $V_{I_{L}}=\{I_{L}\left(i\right)\}$, containing
only one element $i$ of $V$ mapped on the index $I_{T}$ by the
invertible map $I_{T}\left(i\right)$. We call $\mathscr{V}$ filtration
of $V$ the sequence of refinements of 
\begin{equation}
\mathcal{V}_{t}=\left\{ V_{I_{t}}\subset V:\,I_{t}\in{\textstyle \prod_{\,k\leq t}}\{1,2,\,...\,,v_{k}\}\right\} ,
\end{equation}
obtained by joining the subsets $V_{I_{t}}\subseteq V$ from the last
layer $V_{I_{T}}$ to the root level, associated with $V$ itself.
We remark that the two filtrations above are defined independently,
apart from the fact that must have the same number of levels $T$,
the definition of pure state is the same given before in Definition
\ref{14(Pure-States)-Let-1}, with the refinements $\mathcal{S}_{t}$
on behalf of $\mathcal{S}_{t}\left(\mathcal{V}\right)$. }
\end{defn}
Using both the filtrations $\mathcal{\mathscr{S}}$ and $\mathcal{\mathscr{V}}$
one can construct a sequence of kernels that ap\-pro\-xi\-ma\-te
$M_{\mu}$ by progressively averaging (actually ``de-averaging'')
over the re\-fi\-ne\-ments, for example, like in the following
nested approximation scheme for the mag\-ne\-ti\-za\-tion kernel
\begin{defn}
\textit{\label{12(Magnetization-Averages)-Let}(Magnetization Averages)
Start from the last layer $t=T$, that is associated directly to the
external nodes, 
\begin{equation}
\Omega\ni m_{I_{T}}^{A_{T}}=\tau_{i\left(I_{T}\right)}^{\alpha\left(A_{T}\right)}.
\end{equation}
For all the other layers we define
\begin{equation}
\left[0,1\right]\ni m_{I_{t}}^{A_{t}}=\frac{1}{|V_{I_{t}}||S_{A_{t}}|}\sum_{i\in V_{I_{t}}}\sum_{\alpha\in S_{A_{t}}}\mu^{\alpha}\tau_{i}^{\alpha},
\end{equation}
down to the root level $t=0$, for which we drop the tree index and
use simply
\begin{equation}
m=\frac{1}{N|S|}\sum_{i\in V}\sum_{\alpha\in S}\mu^{\alpha}\tau_{i}^{\alpha}=\frac{1}{N}\sum_{i\in V}\langle\,\boldsymbol{\sigma}_{i}\rangle_{\mu}.
\end{equation}
Starting from the Magnetization Averages we define the parameters
\begin{equation}
\delta m_{I_{t}}^{A_{t}}=m_{I_{t}}^{A_{t}}-m_{I_{t-1}}^{A_{t-1}}
\end{equation}
that indicate the fluctuation of the magnetization respect to the
average of the cell at the level $t-1$. For the the root level $t=0$
we simply write $\delta m=m$.} \textit{In this representation the
measure $\mu$ is obtained from the initial condition
\begin{equation}
\eta_{I_{T}}^{A_{T}}\left(\sigma_{I_{T}}\right)={\textstyle \frac{1}{2}}{\textstyle \left(1+\sigma_{I_{T}}\sum_{t=0}^{T}\,\delta m_{I_{t}}^{A_{t}}\right)}
\end{equation}
and then applying the recursive formula
\begin{equation}
\eta_{I_{t-1}}^{A_{t-1}}(\,\sigma_{V_{I_{t-1}}})=\frac{1}{n_{t}}\sum_{\alpha_{t}=1}^{n_{t}}\prod_{i_{t}=1}^{v_{t}}\,\eta_{I_{t}}^{A_{t}}(\,\sigma_{V_{I_{t}}})
\end{equation}
to the last level, the measure $\eta$ that we would like to adapt
to $\mu$, 
\begin{equation}
\eta\left(\sigma_{V}\right)=\frac{1}{n_{1}}\sum_{\alpha_{1}=1}^{n_{1}}\prod_{i_{1}=1}^{v_{1}}\,\eta_{\,i_{1}}^{\alpha_{1}}\left(\sigma_{V_{i_{1}}}\right).
\end{equation}
We remark that, by construction, the averages of any $\delta m$ respect
to the indexes $\alpha_{t}$ and $i_{t}$ is zero, and any averaged
magnetization can be reconstructed from the increments by summing
back them together. The scheme is shown using kernel representation
in Figures \ref{fig3.1-1-1-1} and \ref{fig3.1-1-1-1-1}. }
\begin{figure}
\begin{raggedright}
\includegraphics[scale=0.155]{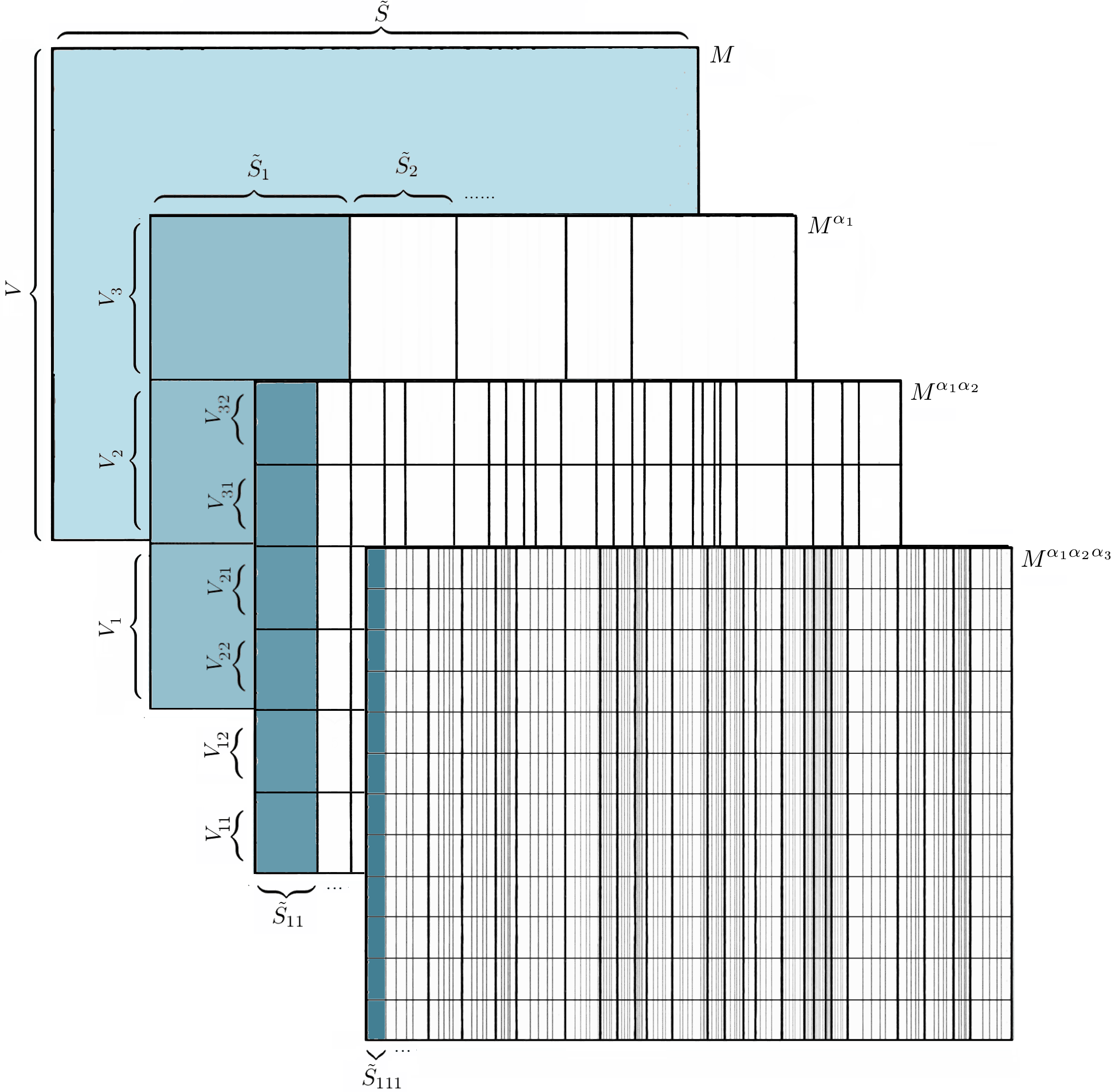}
\par\end{raggedright}

~\caption{\label{fig3.1-1-1-1} Kernel representation of the filtration process
in Definition \ref{12(Magnetization-Averages)-Let} for a 3-RSB system
($L=3$), and partition parameters $n_{0}=5$, $n_{1}=5$ and $v_{0}=3$,
$v_{1}=2$. The measure $\mu$ is the same of Figure \ref{replicated}.
The vertical lines highlight the pure states of each layer $M$, $M^{\alpha_{1}}$
and $M^{\alpha_{1}\alpha_{2}}$ of the kernel $M_{\mu}$. The last
kernel is $M_{\mu}$ itself. The filtration has been chosen to match
that of Figure \ref{fig4.1-2}.}
\end{figure}
\begin{figure}
\begin{centering}
\includegraphics[scale=0.175]{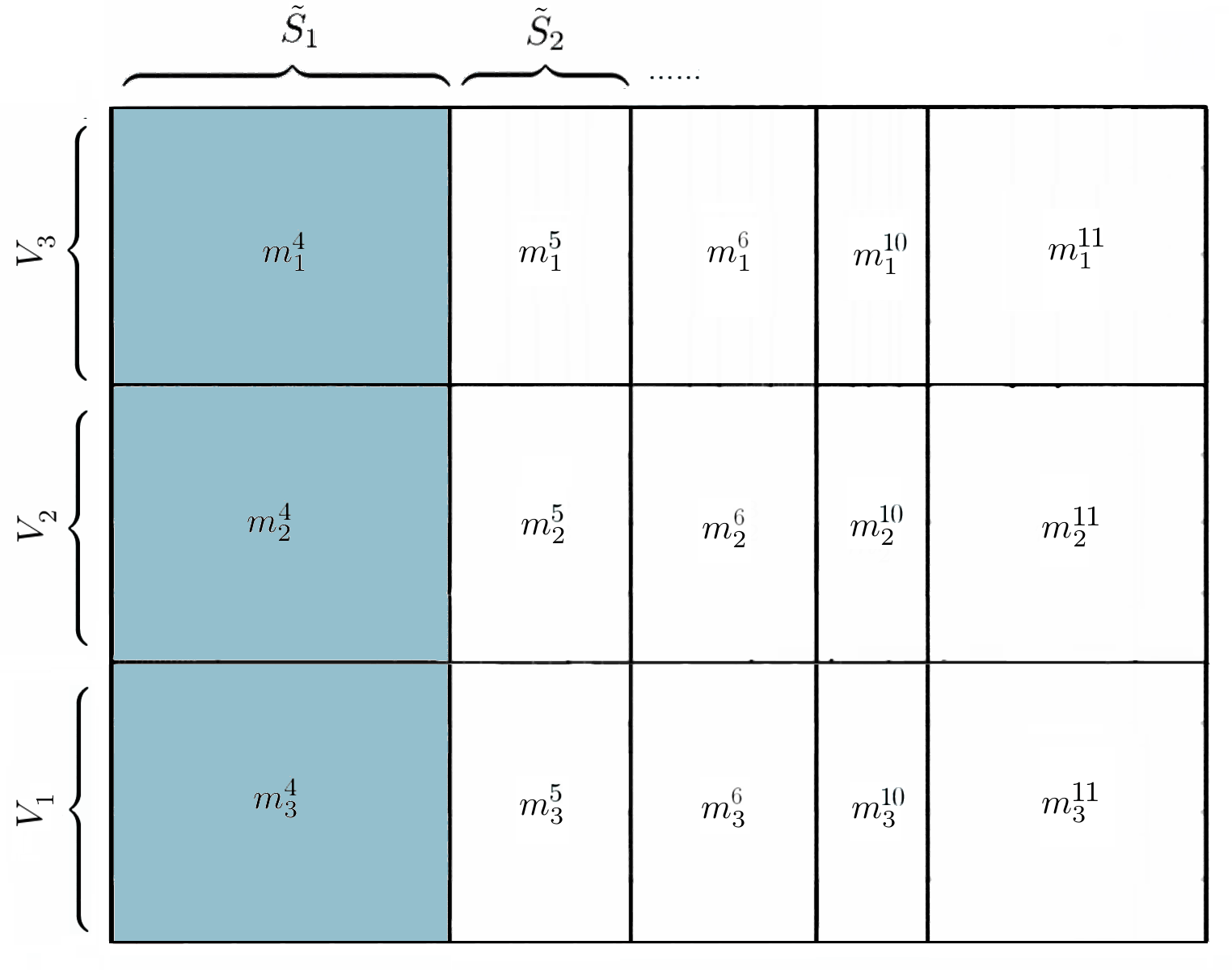}
\par\end{centering}

~\caption{\label{fig3.1-1-1-1-1} Detail of the kernel filtration shown in Figure
(\ref{fig3.1-1-1-1}), associated to the first effective level of
the scheme, that is 1RSB. Each block of the kernel has constant magnetization
$m_{i_{1}}^{\alpha_{1}}$ as given in Definition \ref{12(Magnetization-Averages)-Let},
with $\alpha_{1}\in\left\{ 4,5,6,10,11\right\} $ and $1\leq i_{1}\leq3$. }
\end{figure}
\textit{ }
\end{defn}
\pagebreak{}

~

~

~

~

~

~

~

~

~

~

~

~

~

\section{The SK model\label{sec:The-Ansatz}}

We can finally apply these concepts to the SK model. We start introducing
the basic quantities of the previous section in the case of Gibbs
measures. Consider a system of $N$ spins, governed by the Hamiltonian
\begin{equation}
H:\,\Omega^{V}\rightarrow\mathbb{R},
\end{equation}
the associated Gibbs measure is
\begin{equation}
\mu\left(\sigma_{V}\right)=\frac{1}{Z}\,e^{-\beta H\left(\sigma_{V}\right)},
\end{equation}
the normalization (partition function) is
\begin{equation}
Z=\sum_{\sigma_{V}\in\Omega^{V}}\,e^{-\beta H\left(\sigma_{V}\right)}=\sum_{\sigma_{V_{1}}\in\Omega^{V_{1}}}\,...\,\sum_{\sigma_{V_{L}}\in\Omega^{V_{L}}}\,e^{-\beta H\left(\sigma_{Q_{L}}\right)}.
\end{equation}
Now, the Hamiltonian of the Sherrington-Kirckpatrick (SK) model \cite{Parisi-2,Bolt}
is 
\begin{equation}
\boldsymbol{H}_{sk}\left(\sigma_{V}\right):=\frac{1}{\sqrt{N}}\sum_{i}\sum_{j<i}\sigma_{i}\boldsymbol{J}_{ij}\sigma_{j}
\end{equation}
with $\boldsymbol{J}$ Gaussian (asymmetric) random matrix with normal
independent entries of uni\-ta\-ry va\-rian\-ce. From now we will
work with a random Hamiltonian instead of a single in\-stan\-ce
of it, so that we don't have to add another index for the disorder
when computing the Gaussian averages $E\left(\,\cdot\,\right)$ (for
which we use this spe\-cial notation). As before, we can define the
partition function 
\begin{equation}
\boldsymbol{Z}=\sum_{\sigma_{V}\in\Omega^{V}}e^{-\beta\boldsymbol{H}_{sk}\left(\sigma_{V}\right)},
\end{equation}
that in this case is a $\boldsymbol{J}-$dependent random quantity. 

\begin{figure}
\begin{raggedleft}
\includegraphics[scale=0.2]{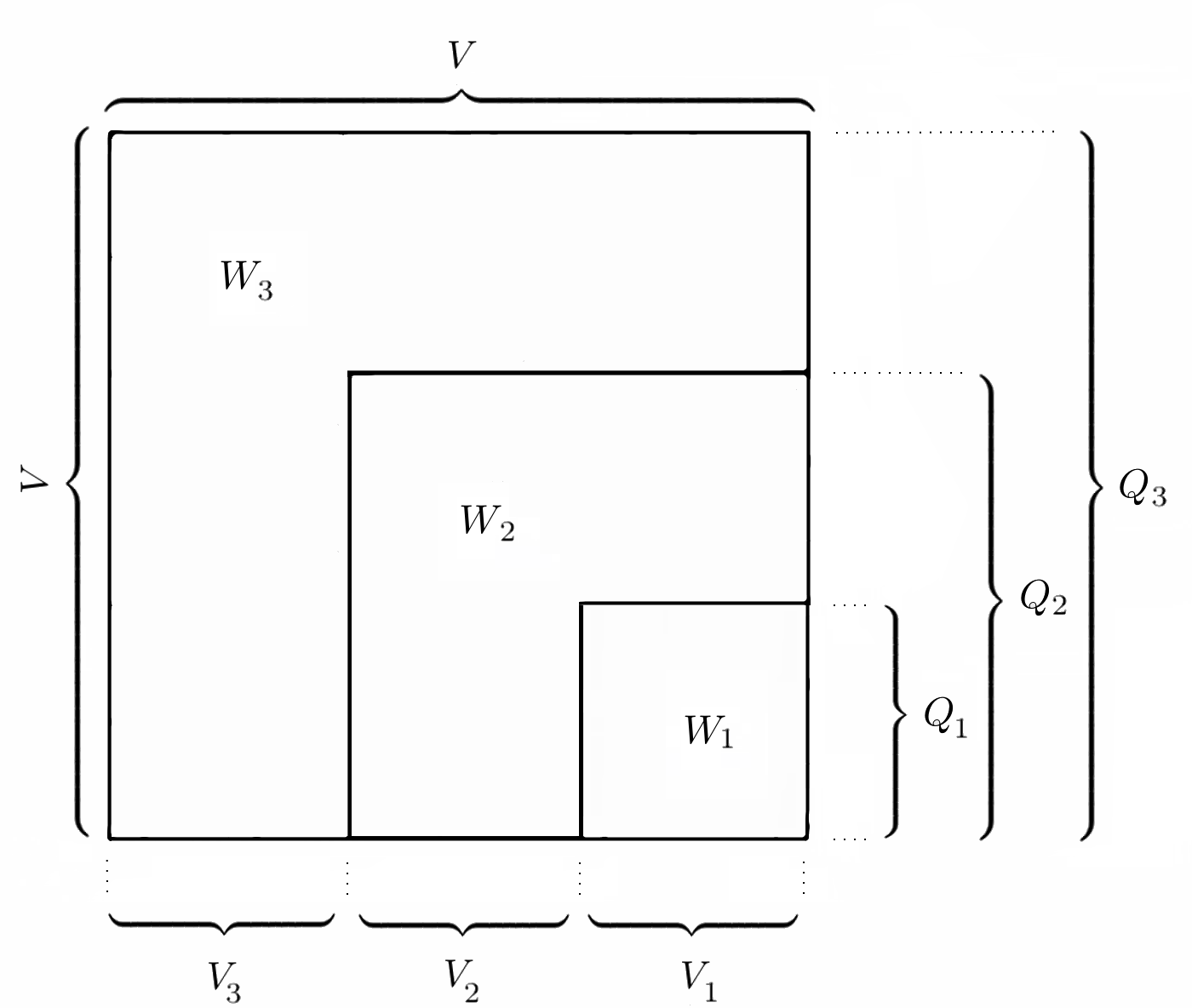}
\par\end{raggedleft}

\raggedleft{}~\caption{\label{fig:Partition-of-}Partition of $W$ induced by $\mathcal{V}$.
Under $\mathcal{V}$ the edges set $W$ is splitted into subsets $W_{\ell}$,
containing all edges with both ends in $Q_{\ell}$ minus those with
both ends in $Q_{\ell-1}$. As predicted in Definition \ref{11(Filtration-of-).},
the contribution to the total energy given by $W_{\ell}$ is adapted
to the spins of $Q_{\ell-1}$.}
\end{figure}

To simplify some of the coming manipulations we will consider the
Asymmetric SK Hamiltonian (ASK), 
\begin{equation}
\boldsymbol{H}\left(\sigma_{V}\right):=\frac{1}{\sqrt{N}}\sum_{i\in V}\sum_{j\in V}\sigma_{i}\boldsymbol{J}_{ij}\sigma_{j},
\end{equation}
because, apart from vanishing finite size corrections, holds 
\begin{equation}
\sqrt{2}\boldsymbol{H}_{sk}\left(\sigma_{V}\right)\stackrel{d}{=}\boldsymbol{H}\left(\sigma_{V}\right)\label{eq:contucci}
\end{equation}
in distribution. Using this definition the temperature is rescaled
by a factor $\sqrt{2}$ respect to the usual Parisi functional. The
functional for the original SK model is recovered from that of ASK
by substituting $\beta$ with $\beta/\sqrt{2}$ .
\begin{lem}
\label{lem:(Layer-States-of}(Layer States of ASK) Given some partition
$\mathcal{V}$ the ASK Hamiltonian can be decomposed according to
Definition \ref{11(Filtration-of-).} as follows 
\begin{equation}
\boldsymbol{H}\left(\sigma_{V}\right)=\sum_{\ell}\boldsymbol{H}_{\ell}\left(\sigma_{Q_{\ell}}\right),
\end{equation}
where the $\boldsymbol{H}_{\ell}$ are the layer Hamiltonians 
\begin{equation}
{\textstyle \boldsymbol{H}_{\ell}\left(\sigma_{Q_{\ell}}\right)}=\frac{1}{\sqrt{N}}\sum_{\left(i,j\right)\in W_{\ell}}\sigma_{i}\boldsymbol{J}_{ij}\sigma_{j}\,.\label{eq:marginalhamiltonian}
\end{equation}
where we introduced the sequence of edges sets

\begin{equation}
W_{\ell}:=Q_{\ell}^{2}\setminus Q_{\ell-1}^{2}.
\end{equation}
~

\noindent In general, we can associate the (random) distributions
\begin{equation}
\boldsymbol{\xi}_{\ell}\left(\sigma_{Q_{\ell}}\right)=\frac{1}{\boldsymbol{Z}_{\ell}\left(\sigma_{Q_{\ell-1}}\right)}e^{-\beta\boldsymbol{H}_{\ell}\left(\sigma_{Q_{\ell}}\right)},
\end{equation}
and the (random) partition functions 
\begin{equation}
\boldsymbol{Z}_{\ell}\left(\sigma_{Q_{\ell-1}}\right)=\sum_{\sigma_{V_{\ell}}\in\Omega^{V_{\ell}}}\,e^{-\beta\boldsymbol{H}_{\ell}\left(\sigma_{Q_{\ell}}\right)}\,.
\end{equation}
\end{lem}
\begin{proof}
The representation of Definition (\ref{11(Filtration-of-).}) for
ASK is as follows. The partition of $V$ is into a number $L$ of
subsets $V_{\ell}$, each of macroscopic size $O\left(N\right)$.
As be\-fore we write everything in terms of the sets $Q_{\ell}$
so that all is controlled by the parameters $|Q_{\ell}|/N=q_{\ell}$.
The sizes of $V_{\ell}$ are then $|V_{\ell}|/N=q_{\ell}-q_{\ell-1}$.
Let 
\begin{equation}
W:=\left\{ \left(i,j\right)\in V^{2}:\,i,j\in V\right\} 
\end{equation}
be the edges set (for the SK this is a fully connected graph). It
is easy to verify that the effect of $\mathcal{V}$ is to produce
a corresponding partition of $W$ into subsets $W_{\ell}$ such that
each $W_{\ell}$ contains all edges with both ends in $Q_{\ell}$
minus those with both ends in $Q_{\ell-1}$ (see Figure \ref{fig:Partition-of-}).
Then we can define a partition of $W$ 
\begin{equation}
\mathcal{W}\left(\mathcal{V}\right)=\left\{ W_{1},W_{2},\,...\,,W_{L}\right\} 
\end{equation}
uniquely defined by the partition $\mathcal{V}$, the sets are $W_{\ell}=Q_{\ell}^{2}\setminus Q_{\ell-1}^{2}.$
The contribution to the total energy given by $W_{\ell}$ is then
\begin{multline}
\boldsymbol{H}_{\ell}\left(\sigma_{Q_{\ell}}\right)=\frac{1}{\sqrt{N}}\sum_{\left(i,j\right)\in W_{\ell}}\sigma_{i}\boldsymbol{J}_{ij}\sigma_{j}=\\
=\frac{1}{\sqrt{N}}\sum_{\left(i,j\right)\in Q_{\ell}^{2}}\sigma_{i}\boldsymbol{J}_{ij}\sigma_{j}-\frac{1}{\sqrt{N}}\sum_{\left(i,j\right)\in Q_{\ell-1}^{2}}\sigma_{i}\boldsymbol{J}_{ij}\sigma_{j}.\label{eq:easy}
\end{multline}

\end{proof}
\begin{figure}
\begin{raggedright}
\includegraphics[scale=0.16]{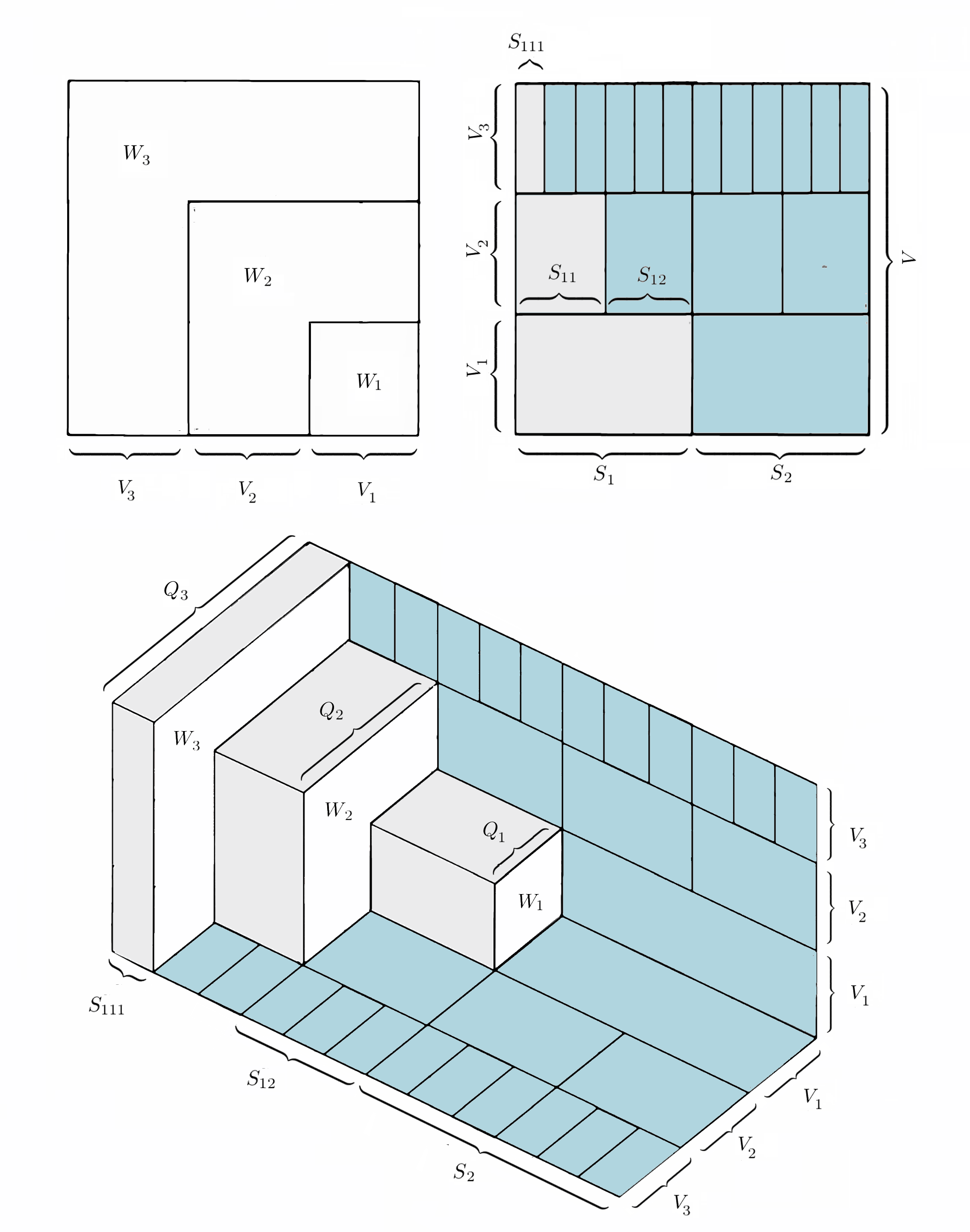}
\par\end{raggedright}

\raggedleft{}~\caption{\label{fig:Partition-of--2} Kernel diagrams of the pair correlations
and their $\mathcal{V}-$partition. The smaller diagrams on top are
the partitions of $S$ and $V^{2}$ described in the captions of the
Figures \ref{fig4.1}, \ref{fig4.1-1} and \ref{fig:Partition-of-}.
The last shows the multi-kernel $\left[0,1\right]^{3}\rightarrow\Omega$
that encodes the tensor $V^{2}S\rightarrow\Omega$ of the pair correlations
$\tau_{i}^{\alpha}\tau_{j}^{\alpha}$, all sub-kernels have been removed
except the first sequence $11...1$ (in gray and white) to highlight
the structure of the pure states.}
\end{figure}

As said in the previous sections, this partition structure is inspired
by the fact that if $\boldsymbol{H}\left(\sigma_{V}\right)$ can be
defined for arbitrary sizes $|V|$ then we should be able to represent
it as the terminal point of the sequence $\sqrt{q_{\ell}}\boldsymbol{H}\left(\sigma_{Q_{\ell}}\right)$.
Notice that the layer Hamiltonians $\boldsymbol{H}_{\ell}$ of Eq.(\ref{eq:marginalhamiltonian})
can in fact be expressed in terms of the difference between two ASK
Hamiltonians, depending on $Q_{\ell}$ and $Q_{\ell-1}$ spins respectively,
ie
\begin{equation}
\boldsymbol{H}_{\ell}\left(\sigma_{Q_{\ell}}\right)=\sqrt{q_{\ell}}\boldsymbol{H}\left(\sigma_{Q_{\ell}}\right)-\sqrt{q_{\ell-1}}\boldsymbol{H}\left(\sigma_{Q_{\ell-1}}\right).
\end{equation}
In this form the layer Hamiltonians allow a better reading of what
we are actually doing, ie reconstructing the system growing it layer
by layer toward a target size $|V|=N$. This can be seen also in the
Figure \ref{fig:Partition-of-}, where a trial partition of the edges
set $W$ is shown. Here the edges $\left(i,j\right)$ are represented
as elements of the square $W=V^{2}$. Remember that the noise of $\boldsymbol{J}_{ij}$
is independent from ed\-ge to ed\-ge, then also between different
$W_{\ell}$. This means that we can average the noise independently
for different $\ell$. 

To better understand the physical meaning it will be convenient to
introduce the \textit{cavity fields} that makes the interface
\begin{equation}
\boldsymbol{h}_{V_{\ell}}\left(\sigma_{Q_{\ell-1}}\right):=\left\{ \boldsymbol{h}_{i}\left(\sigma_{Q_{\ell-1}}\right)\in\mathbb{R}:\,i\in V_{\ell}\right\} ,
\end{equation}
with local components given by
\begin{equation}
\boldsymbol{h}_{i}\left(\sigma_{Q_{\ell-1}}\right):=\frac{1}{\sqrt{|Q_{\ell-1}|}}\sum_{j\in Q_{\ell-1}}\boldsymbol{J}_{ij}\sigma_{j},
\end{equation}
and same $\boldsymbol{J}$ used for the Hamiltonian. Then, the layer
can be rewritten as
\begin{equation}
\boldsymbol{H}_{\ell}\left(\sigma_{Q_{\ell}}\right)=\sqrt{q_{\ell}-q_{\ell-1}}\boldsymbol{H}\left(\sigma_{V_{\ell}}\right)+\sqrt{2q_{\ell-1}}\sigma_{V_{\ell}}\boldsymbol{h}_{V_{\ell}}\left(\sigma_{Q_{\ell-1}}\right),\label{MARGINALSSSSSHAM}
\end{equation}

\noindent where the self-interaction is simply a smaller ASK Hamiltonian
$\boldsymbol{H}\left(\sigma_{V_{\ell}}\right)$, while the contribution
from the interface is mediated by the cavity fields 
\begin{equation}
\bar{\boldsymbol{H}}_{\ell}\left[\sigma_{V_{\ell}},\boldsymbol{h}_{V_{\ell}}\left(\sigma_{Q_{\ell-1}}\right)\right]:=\sigma_{V_{\ell}}\boldsymbol{h}_{V_{\ell}}\left(\sigma_{Q_{\ell-1}}\right)
\end{equation}
and match the Hamiltonian of an Asymmetric Bipartite SK model (ABSK,
see \cite{Barra}) at slightly shifted temperature, and with ratio
between the group sizes that shrinks as $L$ increases. Introducing
the auxiliary temperature variables
\begin{equation}
\beta_{\ell}^{*}:=\beta\sqrt{q_{\ell}-q_{\ell-1}},\ \ \ \beta_{\ell}:=\beta\sqrt{2q_{\ell-1}}
\end{equation}
we arrive to the expression
\begin{equation}
\beta\boldsymbol{H}_{\ell}\left(\sigma_{Q_{\ell}}\right)=\beta_{\ell}^{*}\boldsymbol{H}\left(\sigma_{V_{\ell}}\right)+\beta_{\ell}\bar{\boldsymbol{H}}_{\ell}\left[\sigma_{V_{\ell}},\boldsymbol{h}_{V_{\ell}}\left(\sigma_{Q_{\ell-1}}\right)\right].\label{MARGINALSSSSSHAM-1}
\end{equation}

From this reformulation one can appreciate the structure of the interactions:
the cavity fields $\boldsymbol{h}_{V_{\ell}}\left(\sigma_{Q_{\ell-1}}\right)$
act as random external fields that depend from the previous level,
and toward which the system tries to align, while the thermal fluctuations
and the Hamiltonian $\boldsymbol{H}\left(\sigma_{V_{\ell}}\right)$
act as perturbations that can introduce more directions for the eigenstates
\cite{Franchini}. 
\begin{lem}
(IO model) For large $N$ and $L$, the Gibbs measure associated to
the\label{lem:(IO-model)-For} layer $\beta\boldsymbol{H}_{\ell}$
converges in distribution to that of the interface $\beta_{\ell}\bar{\boldsymbol{H}}_{\ell}$
.\end{lem}
\begin{proof}
First we notice that the term $\boldsymbol{H}\left(\sigma_{V_{\ell}}\right)$
is multiplied by $\beta\sqrt{q_{\ell}-q_{\ell-1}}$ and its role become
less important as $L$ increases. Then, for any finite temperature
$\beta$ we can make $N$ and $L$ large enough to have a sequence
for which $\beta_{\ell}^{*}<\beta_{c}$ at any $\ell$, and it is
established since \cite{Ruelle} and \cite{Frolich} by second moment
methods that in the high temperature regime the annealed averages
match the quenched ones. 
\end{proof}
\noindent The layers can be approximated in distribution by the (random)
relative weights of the interface only (hereafter IO model), 
\begin{equation}
\bar{\boldsymbol{\xi}}_{\ell}\left(\sigma_{Q_{\ell}}\right):=\frac{1}{\boldsymbol{\bar{Z}}_{\ell}\left(\sigma_{Q_{\ell-1}}\right)}e^{-\beta_{\ell}\bar{\boldsymbol{H}}_{\ell}\left[\sigma_{V_{\ell}},\boldsymbol{h}_{V_{\ell}}\left(\sigma_{Q_{\ell-1}}\right)\right]}
\end{equation}
with (random) partition function given by 
\begin{multline}
\boldsymbol{\bar{Z}}_{\ell}\left(\sigma_{Q_{\ell-1}}\right):=\,\sum_{\sigma_{V_{\ell}}\in\Omega^{V_{\ell}}}\,e^{-\beta_{\ell}\bar{\boldsymbol{H}}_{\ell}\left[\sigma_{V\ell},\boldsymbol{h}_{V_{\ell}}\left(\sigma_{Q_{\ell-1}}\right)\right]}=\\
=\sum_{\sigma_{V_{\ell}}\in\Omega^{V_{\ell}}}\,e^{-\beta_{\ell}\sigma_{V_{\ell}}\boldsymbol{h}_{V_{\ell}}\left(\sigma_{Q_{\ell-1}}\right)}=\\
=\prod_{i_{\ell}\in V_{\ell}}2\cosh\left[\beta_{\ell}\boldsymbol{h}_{i_{\ell}}\left(\sigma_{Q_{\ell-1}}\right)\right].\label{eq:IOmodel}
\end{multline}
The interface is simply a group of independent spins coupled to a
field, that is ad\-ap\-ted to the previous layers but does not depend
on the one on which acts. To make some stronger statement it will
be convenient to introduce the (random) spin vector representing the
direction of the external field $\boldsymbol{h}_{V_{\ell}}$, 
\begin{equation}
\boldsymbol{\omega}_{V_{\ell}}\left(\sigma_{Q_{\ell-1}}\right):=\left\{ \boldsymbol{\omega}_{i_{\ell}}\left(\sigma_{Q_{\ell-1}}\right)\in\Omega:\,i_{\ell}\in V_{\ell}\right\} ,
\end{equation}
 that we call \textit{master direction}, its components are defined
as follows:
\begin{equation}
\boldsymbol{\omega}_{i_{\ell}}\left(\sigma_{Q_{\ell-1}}\right):=\boldsymbol{h}_{i_{\ell}}\left(\sigma_{Q_{\ell-1}}\right)/\left|\boldsymbol{h}_{i_{\ell}}\left(\sigma_{Q_{\ell-1}}\right)\right|.\label{eq:rttttrtrationz}
\end{equation}
Notice that due to parity of the $\cosh$ function the partition function
$\boldsymbol{\bar{Z}}_{\ell}$ does not depend on $\boldsymbol{\omega}_{V_{\ell}}$
but only on the projections of $\boldsymbol{\omega}_{i_{\ell}}$ on
the local fields $\boldsymbol{h}_{i_{\ell}}$. This vector has all
positive entries $\left|\boldsymbol{h}_{i_{\ell}}\right|$ and can
be represented by the Hadamard product between $\boldsymbol{\omega}_{V_{\ell}}$
and $\boldsymbol{h}_{V_{\ell}}$. Let also introduce the local overlap
of the $i_{\ell}-$th spin with the direction of the external field
\begin{multline}
\boldsymbol{m}_{i_{\ell}}\left(\sigma_{Q_{\ell-1}}\right):=\langle\,\sigma_{i_{\ell}}\rangle_{\bar{\boldsymbol{\xi}}_{\ell}\left(\sigma_{Q_{\ell-1}}\right)}\boldsymbol{\omega}_{i_{\ell}}\left(\sigma_{Q_{\ell-1}}\right)=\\
=\left|\,\tanh\left[\beta_{\ell}\boldsymbol{h}_{i_{\ell}}\left(\sigma_{Q_{\ell-1}}\right)\right]\right|\in\left[0,1\right]\label{eq:dsfsffr}
\end{multline}
this parameter is a measure of how much the spin $\sigma_{i_{\ell}}$
is binded to the direction of the external field, and depends on the
amplitude of $\boldsymbol{h}_{i_{\ell}}$. 

The parameter $\boldsymbol{m}_{i_{\ell}}$ is an analogue of the local
magnetization, is also related to the local overlap by the formula
\begin{equation}
\boldsymbol{m}_{i_{\ell}}\left(\sigma_{Q_{\ell-1}}\right)=\sqrt{\langle\,\sigma_{i_{\ell}}\tau_{i_{\ell}}\,\rangle_{\bar{\boldsymbol{\xi}}_{\ell}\left(\sigma_{Q_{\ell-1}}\right)\otimes\,\bar{\boldsymbol{\xi}}_{\ell}\left(\sigma_{Q_{\ell-1}}\right)}},
\end{equation}
and can be used as local order parameter. If the amplitude $\left|\boldsymbol{h}_{i_{\ell}}\right|$
(or $\beta$) is large, the spin will be forced to align with the
field and $\boldsymbol{m}_{i_{\ell}}\rightarrow1$. On the contrary,
when $\boldsymbol{h}_{i_{\ell}}$ is small, or $\beta$ is small,
then $\boldsymbol{m}_{i_{\ell}}\rightarrow0$ as the spin disentangles
from the direction $\boldsymbol{\omega}_{i_{\ell}}$. The fluctuations
of the interfaces can be characterized in detail by studying the kernel
of $\boldsymbol{\sigma}_{V_{\ell}}^{*}:=\sigma_{V_{\ell}}\circ\boldsymbol{\omega}_{V_{\ell}}$
Hadamard product between $\sigma_{V_{\ell}}$ and $\boldsymbol{\omega}_{V_{\ell}}$,
the components
\begin{equation}
\boldsymbol{\sigma}_{i_{\ell}}^{*}\left(\sigma_{Q_{\ell-1}}\right):=\sigma_{i_{\ell}}\,\boldsymbol{\omega}_{i_{\ell}}\left(\sigma_{Q_{\ell-1}}\right).
\end{equation}
Notice that the scalar product of spin states (overlap) equals the
magnetization of their Hadamard product, also notice that the Hadamard
product of a group of spin states by a common master spin state is
a theta-map of the kind described in Section \ref{sec:Kernel-representation},
and does not change the overlap between the states in the transformed
group. Since we are now analyzing the behavior inside a given layer
state, we can drop the dependence on previous layers. First notice
that the layer Hamiltonian can be written in terms of the $\boldsymbol{\sigma}_{V_{\ell}}^{*}$
vector 
\begin{equation}
\sum_{i\in V_{\ell}}\,\sigma_{i}\boldsymbol{h}_{i}=\sum_{i\in V_{\ell}}\,\sigma_{i}\left(\boldsymbol{\omega}_{i}^{2}\right)\boldsymbol{h}_{i}=\sum_{i\in V_{\ell}}\,\left(\sigma_{i}\boldsymbol{\omega}_{i}\right)\left(\boldsymbol{\omega}_{i}\boldsymbol{h}_{i}\right)=\sum_{i\in V_{\ell}}\,\boldsymbol{\sigma}_{i}^{*}\left|\boldsymbol{h}_{i}\right|,
\end{equation}
the relative orientations of $\boldsymbol{\sigma}_{V}^{*}$ and $\sigma_{V}$
are randomized by the multiplication with the random direction $\boldsymbol{\omega}_{i}$,
but are not independent, for example their overlap $\boldsymbol{\sigma}_{V}^{*}\sigma_{V}$
is the total magnetization of the master direction $\boldsymbol{\omega}_{V}$.
Since $\sigma_{V}=\boldsymbol{\sigma}_{V_{\ell}}^{*}\circ\boldsymbol{\omega}_{V}$,
for a general non-random function $f$ it will be convenient to introduce
the associated random function 
\begin{equation}
\boldsymbol{f}^{*}\left(\sigma_{V}\right):=f\left(\boldsymbol{\sigma}_{V}^{*}\right)=f\left(\sigma_{V}\circ\boldsymbol{\omega}_{V}\right),
\end{equation}
then we can simplify the notation by rewriting $f\left(\sigma_{V}\right)$
as follows 
\begin{equation}
f\left(\sigma_{V}\right)=f\left(\boldsymbol{\sigma}_{V_{\ell}}^{*}\circ\boldsymbol{\omega}_{V}\right)=\boldsymbol{f}^{*}\left(\boldsymbol{\sigma}_{V_{\ell}}^{*}\right)
\end{equation}
and recast the spin variables on which the average is applied: 
\begin{multline}
\langle f\left(\boldsymbol{\sigma}_{V_{\ell}}\right)\rangle_{\xi_{\ell}}=\frac{\sum_{\sigma_{V_{\ell}}\in\Omega^{V_{\ell}}}e^{-\beta_{\ell}\sum_{i\in V_{\ell}}\,\boldsymbol{\sigma}_{i}^{*}\left|\boldsymbol{h}_{i}\right|}f\left(\sigma_{V_{\ell}}\right)}{\sum_{\sigma_{V_{\ell}}\in\Omega^{V_{\ell}}}e^{-\beta_{\ell}\sum_{i\in V_{\ell}}\,\boldsymbol{\sigma}_{i}^{*}\left|\boldsymbol{h}_{i}\right|}}=\\
=\frac{\sum_{\sigma_{V_{\ell}}\in\Omega^{V_{\ell}}}e^{-\beta_{\ell}\sum_{i\in V_{\ell}}\,\sigma_{i}\left|\boldsymbol{h}_{i}\right|}\boldsymbol{f}^{*}\left(\sigma_{V_{\ell}}\right)}{\sum_{\sigma_{V_{\ell}}\in\Omega^{V}}e^{-\beta_{\ell}\sum_{i\in V_{\ell}}\,\sigma_{i}\left|\boldsymbol{h}_{i}\right|}}\label{eq:dfgdsgdd-1}
\end{multline}
ie, we have changed the sum from $\boldsymbol{\sigma}_{V_{\ell}}^{*}$
to $\sigma_{V_{\ell}}$, remembering that the two vectors are uniquely
linked and both span the same space. 

There is still a little technical difficulty in that $\left|\boldsymbol{h}_{i}\right|$
are different from site to site. We can overcome this problem in various
ways, for example by separating the average from the residual fluctuations
of the field: 
\begin{equation}
\psi_{\ell}:=\frac{1}{|V_{\ell}|}\sum_{i\in V_{\ell}}\left|\boldsymbol{h}_{i}\right|,\ \ \ \boldsymbol{\varphi}_{i}:=\left|\boldsymbol{h}_{i}\right|-\frac{1}{|V_{\ell}|}\sum_{i\in V_{\ell}}\left|\boldsymbol{h}_{i}\right|.\label{eq:fields}
\end{equation}
Notice that for $N\rightarrow\infty$ the average amplitude $\psi_{\ell}$
converges to a deterministic constant that only depends on the input
from the previous layers,

\begin{equation}
\sum_{i\in V_{\ell}}\,\sigma_{i}\left|\boldsymbol{h}_{i}\right|=\psi_{\ell}\,M\left(\sigma_{V_{\ell}}\right)+\sigma_{V_{\ell}}\cdot\boldsymbol{\varphi}_{V_{\ell}},
\end{equation}
where $M\left(\sigma_{V}\right)$ denotes the total magnetization
of $\sigma_{V}$ (that should not be confused with the kernel of the
previous sections). As we shall see in short, this expression already
al\-lows to establish the connection with the Random Energy Model.

An alternative way is to realize that the layer Hamiltonian $\boldsymbol{H}_{\ell}$
is essentially a Num\-ber Partitioning Problem (NPP) on the number
sequence $\left|\boldsymbol{h}_{i}\right|$: the NPR would ac\-tually
be $\left|\boldsymbol{H}_{\ell}\right|$ but the function $\boldsymbol{H}_{\ell}^{2}$
is the same for both models. Then, following Borgs, Chayes and Pittel
\cite{BCP theorem,BCP 2} (see also \cite{Bauke Frnz Mertens} for
an informal discussion), we re\-order the spins of $V_{\ell}$ such
that the $\left|\boldsymbol{h}_{i}\right|$ amplitudes form a non-decreasing
sequence in $i$, ie such that $\left|\boldsymbol{h}_{i+1}\right|\geq\left|\boldsymbol{h}_{i}\right|$,
this is possible because the fields $\boldsymbol{h}_{i}$ are extracted
independently for different index and there is no space structure
to preserve. Then, we can further divide the layers into so\-me large
number $L'$ of \textit{sub-layers} $V_{\ell\ell'}$, of equal volume,
marked by $\ell'$: inside these sub-layers the fluctuations are bounded
\begin{equation}
\psi_{\ell\ell'}^{-}\leq\left|\boldsymbol{h}_{i}\right|\leq\psi_{\ell\ell'}^{+},\ \ \delta_{\ell\ell}:=|\psi_{\ell\ell'}^{+}-\psi_{\ell\ell'}^{-}|\leq c_{0}/L'
\end{equation}
almost everywhere by some constant $c_{0}/L'$ with finite $c_{0}$,
so that the fluctuations goes to zero for large $L'$ and can be neglected.
The \textit{sub-layers} converge to 
\begin{equation}
\bar{\boldsymbol{H}}_{\ell\ell'}\left(\sigma_{V_{\ell\ell'}},\boldsymbol{h}_{V_{\ell\ell'}}\right):=\psi_{\ell\ell'}\,M\left(\sigma_{V_{\ell\ell'}}\right),
\end{equation}
where $\psi_{\ell\ell'}$ is the average of $\left|\boldsymbol{h}_{i}\right|$
inside $V_{\ell\ell'}$, then 
\begin{equation}
\bar{\boldsymbol{H}}_{\ell}\left[\,\sigma_{V_{\ell}},\boldsymbol{h}_{V_{\ell}}\left(\sigma_{Q_{\ell-1}}\right)\right]\rightarrow\sum_{\ell'\leq L'}\psi_{\ell\ell'}\left[\boldsymbol{h}_{V_{\ell}}\left(\sigma_{Q_{\ell-1}}\right)\right]M\left(\sigma_{V_{\ell\ell'}}\right),
\end{equation}
notice that this last representation allows to study the sub-layer
in terms of the magnetization eigenstates only, although this is done
at the price of introducing a new level of partition. 

\pagebreak{}

\section{Thermodynamics of the interface\label{sec:Thermodynamics-of-the}}

To analyze the sub-layers it will be convenient to further simplify
the notation and drop the $\ell\ell'$ index for this section. Let
$\xi$ the sub-layer measure for some fixed pair $\ell$ and $\ell'$,
then, the average of a function can be expressed as: 
\begin{equation}
\langle f\left(\boldsymbol{\sigma}_{V}\right)\rangle_{\xi}=\frac{\sum_{\,\sigma_{V}\in\Omega^{V}}\,e^{-\beta\psi M\left(\sigma_{V}\right)-\beta\boldsymbol{\sigma}_{V}\cdot\boldsymbol{\varphi}_{V}}\boldsymbol{f}^{*}\left(\sigma_{V}\right)}{\sum_{\,\sigma_{V}\in\Omega^{V}}\,e^{-\beta\psi M\left(\boldsymbol{\sigma}_{V}\right)-\beta\boldsymbol{\sigma}_{V}\cdot\boldsymbol{\varphi}_{V}}}.
\end{equation}
We first do the average according to $M$. Let introduce the set of
magnetization eigenstates with given eigenvalue $M$ (see Figure \ref{figANSAZ}):
\begin{equation}
\Omega\left(M,N\right):=\left\{ \sigma_{V}\in\Omega^{V}:\,M\left(\sigma_{V}\right)=M\right\} ,\label{eq:MAG_EIGENSTAT}
\end{equation}
In what follows it will be convenient to also define a simplified
notation $\Omega\left(m\right)$, to indicate set of the magnetization
eigenstates with given eigenvalue $M=\left\lfloor mN\right\rfloor $,
where $\left\lfloor mN\right\rfloor $ is the lower integer part of
$mN$ and $m\in\left[-1,1\right]$.
\begin{lem}
Let $\gamma$ be the Gibbs measure associated to the Hamiltonian $M\left(\sigma_{V}\right)$:
\begin{equation}
\gamma\left(\sigma_{V}\right):=\frac{1}{Z}\,e^{-\beta\psi M\left(\sigma_{V}\right)},
\end{equation}
the partition function is simply $\log Z=N\log2\cosh\left(\beta\psi\right)$,
for large systems the average of some test function $f:\Omega^{V}\rightarrow\mathbb{R}$
respect to $\gamma$ converges to 
\begin{equation}
\langle f\left(\boldsymbol{\sigma}_{V}\right)\rangle_{\gamma}=\langle\,f\left(\boldsymbol{\sigma}_{V}\right)\rangle_{\Omega\left(m\right)},
\end{equation}
the order parameter $m:=\tanh\left(\psi\beta\right)$ is the limit
magnetiza\-tion at which the test function $f$ is sampled.\end{lem}
\begin{proof}
The set $\Omega\left(m\right)$ can be studied using Large Deviations
Theory (LDT), even at the sample-path LDT level. 
\begin{figure}[p]
\raggedright{}\includegraphics[scale=0.22]{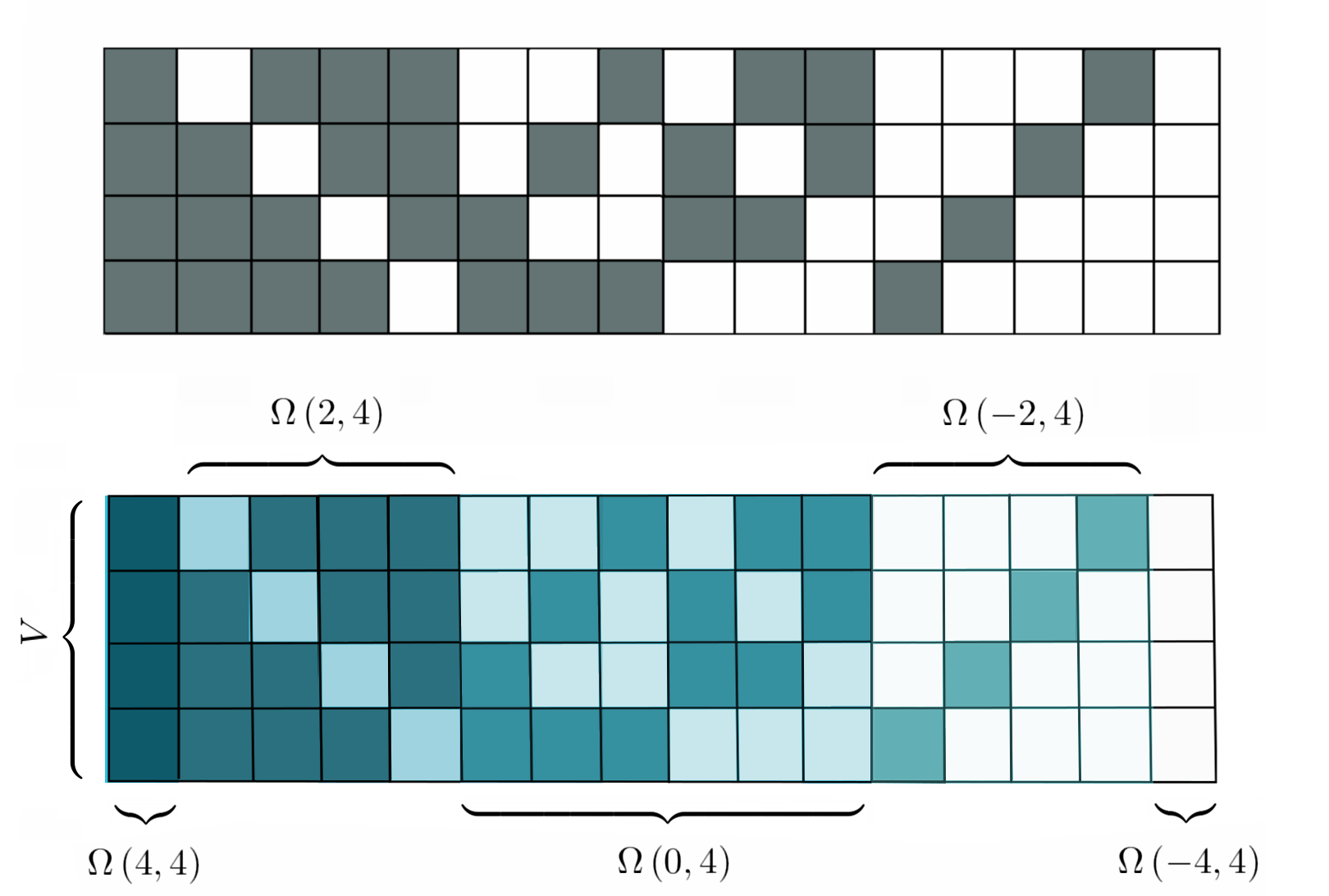} \caption{\label{figANSAZ} Example of a partition of $\Omega^{V}$, with $N=4$,
into the magnetization ei\-gen\-sta\-tes $\Omega\left(M,N\right)$
of Eq. (\ref{eq:MAG_EIGENSTAT}). The spin states are organized by
decreasing total mag\-ne\-ti\-za\-tion. The sub-kernels associated
to groups of spins with given magnetization $M$ are highlighted in
various shades. Let $X$ be the number of spin up in a given spin
state, then, the kernels with fixed $M$ are equivalent to self-avoiding
lattice gases of $X=N/2-M/2$ particles on a lattice of size $N$.}
\end{figure}
 For example, following the methods presented in the proof sections
of \cite{FranchiniUrns}, ie by some simple applications of the Varadhan
Integral Lemma and other standard LDT theory tools, one can compute
$|\Omega\left(m\right)|$ and find that is proportional to $\exp\left[N\phi\left(m\right)\right]$
with $\phi$ convex function of $m$. Given that these methods are
well known we only give the essential features. 

~

~

~

The average according to $\gamma$ can be expressed in terms of the
magnetizations eigenstates of as follows:
\begin{equation}
\langle f\left(\boldsymbol{\sigma}_{V}\right)\rangle_{\gamma}=\frac{\sum_{M}\left|\Omega\left(M,N\right)\right|e^{-\beta M}\,\langle f\left(\boldsymbol{\sigma}_{V}\right)\rangle_{\Omega\left(M,N\right)}}{\sum_{M}\left|\Omega\left(M,N\right)\right|e^{-\beta M}}.
\end{equation}
In the last formula we introduced a braket notation for the uniform
average on the eigenstates of magnetization: for integer $M$ we write
\begin{equation}
\langle f\left(\boldsymbol{\sigma}_{V}\right)\rangle_{\Omega\left(M,N\right)}:=\frac{1}{|\Omega\left(M,N\right)|}\sum_{\sigma_{V}\in\Omega\left(M,N\right)}f\left(\boldsymbol{\sigma}_{V}\right),
\end{equation}
that simplifies to $\langle f\left(\boldsymbol{\sigma}_{V}\right)\rangle_{\Omega\left(m\right)}$
in case $M=\left\lfloor mN\right\rfloor $. In this case is also possible
to re\-pre\-sent the average in integral form: for any test function
$f$ 
\begin{equation}
\sum_{\sigma_{V}\in\Omega^{V}}e^{-\beta M\left(\sigma_{V}\right)}f\left(\sigma_{V}\right)\propto\int_{m\in\left[-1,1\right]}dm\,e^{-Np\left(\beta,m\right)}\langle\,f\left(\boldsymbol{\sigma}_{V}\right)\rangle_{\Omega\left(m\right)}.
\end{equation}

The value at which $m$ concentrates can be the computed using a saddle
point me\-thod applied to the pressure functional $p\left(\beta,m\right):=\beta m-\phi\left(m\right)$
with entropy functional $\phi\left(m\right)$ given by the formula
\begin{multline}
\phi\left(m\right)={\textstyle \left(\frac{1+m}{2}\right)\log\left(\frac{1+m}{2}\right)+\left(\frac{1-m}{2}\right)\log\left(\frac{1-m}{2}\right)}=\\
{\textstyle =-\log2+\frac{1}{2}\log\left(1-m^{2}\right)+\frac{m}{2}\log\left(\frac{1+m}{1-m}\right).}\label{fdfdff-1}
\end{multline}
Follows $\partial_{m}\phi\left(m\right)=\tanh^{-1}\left(m\right)$,
and $\partial_{m}^{2}\phi\left(m\right)=1/\left(1-m^{2}\right)$.
Then we compute the ave\-ra\-ge magnetization $m\left(\beta\right)$
by putting the derivative of the pressure to zero, $\partial_{m}p\left(\beta,m\right)=0$,
that is equivalent to impose $\partial_{m}\phi\left(m\right)=\beta$.
In the end one obtains that $m\left(\beta\right)=\tanh\left(\beta\right)$
and finds
\begin{equation}
\sum_{\sigma_{V}\in\Omega^{V}}e^{-\beta M\left(\sigma_{V}\right)}f\left(\sigma_{V}\right)\propto\langle\,f\left(\boldsymbol{\sigma}_{V}\right)\rangle_{\Omega\left(m\left(\beta\right)\right)},
\end{equation}
we remark once again that the average of $\boldsymbol{\sigma}_{V}$
on the states $\Omega\left(m\right)$ is taken with equal weights
for each magnetization eigenstate. 
\end{proof}
\noindent Then, in the limit of large $N$ the sub-layer average is
\begin{equation}
\langle f\left(\boldsymbol{\sigma}_{V}\right)\rangle_{\xi}=\frac{\sum_{\,\sigma_{V}\in\Omega\left(m\right)}e^{-\beta\boldsymbol{\sigma}_{V}\cdot\boldsymbol{\varphi}_{V}}\boldsymbol{f}^{*}\left(\boldsymbol{\sigma}_{V}\right)}{\sum_{\,\sigma_{V}\in\Omega\left(m\right)}e^{-\beta\boldsymbol{\sigma}_{V}\cdot\boldsymbol{\varphi}_{V}}},
\end{equation}
the next lemma we show that at very low temperature the sub-layer
fluctuations converge in distribution to the Random Energy Model of
Derrida (REM, see also \cite{Bovier-Kurkova-1,Panchenko,Bolt} for
reviews).
\begin{lem}
(Random Energy Model\label{lem:(Random-Energy-Model-1}) Let $\boldsymbol{\varphi}_{V}\in\mathbb{R}^{V}$
be some random vector with $N$ inde\-pen\-dent entries of variance
$\delta$ as defined in Eq. (\ref{eq:fields}), and let $\Omega\left(m\right)$
be the set of magnetization eigenstates of eigenvalue $\left\lfloor mN\right\rfloor $,
\begin{equation}
m:=\tanh\left(\psi\beta\right).
\end{equation}
Let $\boldsymbol{\xi}:\Omega^{V}\rightarrow\left[0,1\right]$ be the
measure of the generic sub-layer 
\begin{equation}
\boldsymbol{\xi}\left(\sigma_{V}\right)=\frac{e^{-\beta\sigma_{V}\cdot\boldsymbol{\varphi}_{V}}}{\sum_{\sigma_{V}\in\Omega\left(m\right)}e^{-\beta\sigma_{V}\cdot\boldsymbol{\varphi}_{V}}}\,I\left(\sigma_{V}\in\Omega\left(m\right)\right),
\end{equation}
at low temperatures, the fluctuations of the sub-layer $\boldsymbol{\xi}$
converge in distribution to a Random Energy Model. \end{lem}
\begin{proof}
For this proof we assume that the support of $\sigma_{V}$ is the
set of magnetization eigenstates for some given magnetization parameter.
Start from 
\begin{equation}
\sum_{i\in V}\sigma_{i}\boldsymbol{\varphi}_{i}=\sum_{i\in V}\boldsymbol{\varphi}_{i}-\sum_{i\in V}\,\left(1-\sigma_{i}\right)\boldsymbol{\varphi}_{i}\label{eq:dfvgd}
\end{equation}
and define the following subset of $V$:
\begin{equation}
X\left(\sigma_{V}\right):=\left\{ j\in V:\,\sigma_{j}=-1\right\} 
\end{equation}
that contains only those sites of $V$ such that $\sigma_{j}$ is
flipped respect to the master direction $\boldsymbol{\omega}_{j}$.
Notice that the size of the set $X$ is fixed, 
\begin{equation}
\frac{1}{N}\,|X\left(\sigma_{V}\right)|=\frac{1-m}{2}=:\epsilon.
\end{equation}
Further noticing that for the term on the right in Eq. (\ref{eq:dfvgd})
holds
\begin{equation}
\sum_{i\in V}\,\left(1-\sigma_{i}\right)\boldsymbol{\varphi}_{i}=\sum_{i\in X\left(\sigma_{V}\right)}2\,\boldsymbol{\varphi}_{i}
\end{equation}
we can define two new cavity variables, a constant offset
\begin{equation}
\boldsymbol{\varphi}\left(1_{V}\right):=\frac{1}{\sqrt{N}}\sum_{i\in V}\boldsymbol{\varphi}_{i}\label{eq:newcavity0}
\end{equation}
and the actual fluctuating term, that is given by
\begin{equation}
\hat{\boldsymbol{\varphi}}\left(\sigma_{V}\right):=\frac{1}{\sqrt{\epsilon N}}\sum_{i\in X\left(\sigma_{V}\right)}\boldsymbol{\varphi}_{i},\label{eq:newcavity}
\end{equation}
then, the fluctuations can be rewritten as
\begin{equation}
\sum_{i\in V}\sigma_{i}\boldsymbol{\varphi}_{i}=\sqrt{N}\,\boldsymbol{\varphi}\left(1_{V}\right)-2\sqrt{\epsilon N}\,\hat{\boldsymbol{\varphi}}\left(\sigma_{V}\right)
\end{equation}
the constant offset $\boldsymbol{\varphi}\left(1_{V}\right)$ cancels
out in the average formula, and it is possible to re\-wri\-te the
average in terms of the $\hat{\boldsymbol{\varphi}}\left(\sigma_{V}\right)$
variables only: we can concentrate on the ac\-tual\-ly fluctuating
component. Notice that by the Central Limit Theo\-rem $\hat{\boldsymbol{\varphi}}\left(\sigma_{V}\right)$
can be approximated in distribution by a sum of Ga\-us\-sian variables
with variance $\delta$, ie 
\begin{equation}
\hat{\boldsymbol{J}}\left(\sigma_{V}\right):=\hat{\boldsymbol{\varphi}}\left(\sigma_{V}\right)/\sqrt{\delta},
\end{equation}
is approximately a Gaussian variable of unitary variance. Here is
the final step: the over\-lap between the flipped sets $X\left(\sigma_{V}\right)$
and $X\left(\tau_{V}\right)$ concentrates on $\epsilon^{2}N$ in
the TL,
\begin{equation}
\lim_{N\rightarrow\infty}\frac{1}{N}\,\langle\,|X\left(\sigma_{V}\right)\cap X\left(\tau_{V}\right)|\,\rangle_{\,\Omega\left(m\right)\otimes\Omega\left(m\right)}=\epsilon^{2},
\end{equation}
and vanishes faster than the size of $X\left(\sigma_{V}\right)$ as
$\beta\rightarrow\infty$, then, the fluctuations $\hat{\boldsymbol{\boldsymbol{J}}}$
become asymptotically independent for each input, like in the REM. \end{proof}
\begin{lem}
\label{lem:Let--be_rem}Let $f_{0}$ be a positive function of the
fluctuations of the cavity fields around the master direction (ground
state): 
\begin{equation}
\beta\boldsymbol{\Delta}^{*}\left(\sigma_{V}\right):=\frac{\beta}{\sqrt{N}}\sum_{i\in V}\boldsymbol{J}_{i}^{*}\left(1-\sigma_{i}\right),
\end{equation}
defined with an independent noise vector $\boldsymbol{J}_{V}^{*}$.
Then, the average of $f_{0}$ according to $\boldsymbol{\xi}$ can
be approximated by a Poisson Point Process of rate $\lambda$: 
\begin{equation}
\langle f_{0}\left[\beta\boldsymbol{\Delta}^{*}\left(\sigma_{V}\right)\right]\rangle_{\boldsymbol{\xi}}^{\lambda}\overset{d}{=}K_{0}\,\langle\,f_{0}[\tilde{\beta}\boldsymbol{\Delta}^{*}\left(\sigma_{V}\right)]{}^{\lambda}\rangle_{\nu},
\end{equation}
where $K_{0}$ is a constant, and $\lambda$ and $\tilde{\beta}$
are deterministic parameters that does not depend on the spins $\sigma_{V}$
of the considered layer (although may still depend from those of the
previous). \end{lem}
\begin{proof}
Form the proof of Lemma \ref{lem:(Random-Energy-Model-1} before,
the average formula is
\begin{equation}
\langle f_{0}\left[\beta\boldsymbol{\Delta}^{*}\left(\sigma_{V}\right)\right]\rangle_{\boldsymbol{\xi}}=\frac{\sum_{\sigma_{V}\in\Omega\left(m\right)}e^{-2\beta\sqrt{\delta\epsilon}\,\hat{\boldsymbol{J}}\left(\sigma_{V}\right)\sqrt{N}}f_{0}[\beta\boldsymbol{\Delta}^{*}\left(\sigma_{V}\right)]}{\sum_{\sigma_{V}\in\Omega\left(m\right)}e^{-2\beta\sqrt{\delta\epsilon}\,\hat{\boldsymbol{J}}\left(\sigma_{V}\right)\sqrt{N}}}.
\end{equation}
and in the low temperature limit we can use the properties of the
REM to study it. It is known that for a REM of (random) Gibbs distribution
\begin{equation}
\boldsymbol{\eta}\left(\sigma_{V}\right)=\frac{e^{-\beta\boldsymbol{J}\left(\sigma_{V}\right)\sqrt{N}}}{\sum_{\,\tau_{V}\in\Omega^{V}}e^{-\beta\boldsymbol{J}\left(\tau_{V}\right)\sqrt{N}}},
\end{equation}
with $\boldsymbol{J}\left(\sigma_{V}\right)$ independent and normally
distributed, holds that for any positive test function $f:\Omega^{V}\rightarrow\mathbb{R}^{+}$
the (random) average is equal to
\begin{equation}
\sum_{\sigma_{V}\in\Omega^{V}}\boldsymbol{\eta}\left(\sigma_{V}\right)f\left(\sigma_{V}\right)\overset{d}{=}K_{0}\langle f\left(\boldsymbol{\sigma}_{V}\right)^{\lambda}\rangle_{\nu}^{1/\lambda}\label{eq:REMMMM}
\end{equation}
with rate parameter $\lambda=\sqrt{\log2}/\beta$ for $\beta>\sqrt{2\log2}$,
and $\lambda=1$ otherwise. 

This result is well known: at low temperatures the weights are proportional
in dis\-tribution to a Poisson Point Process (PPP) of rate $\lambda$,
due to concentration of the mea\-sure on the states with lowest energy,
and by applying the fundamental averaging property of PPP \cite{Panchenko,Bolt}
(see also Little Theorem of \cite{Parisi mezard}) Eq.(\ref{eq:REMMMM})
follows. Above the threshold the sampling of the test function is
dense and the ave\-rage is unaffected. We can adapt the formula of
Eq. (\ref{eq:REMMMM}) to our case using the sca\-ling pro\-perties
of REM: first rescale the number of spins to take into account the
size of $\Omega\left(m\right)$: we define $K:=N/N_{c}$ with $N_{c}:=\log2/\phi\left(m\right)$,
then rescale the fluc\-tua\-tions of the cavity fields from $N$
to $K$ spins 
\begin{multline}
\boldsymbol{\Delta}^{*}\left(\sigma_{V}\right):=\frac{1}{\sqrt{N}}\sum_{i\in V}\boldsymbol{J}_{i}^{*}-\frac{1}{\sqrt{N}}\sum_{i\in V}\boldsymbol{J}_{i}^{*}\sigma_{i}=\\
=\frac{2}{\sqrt{N}}\sum_{i\in X\left(\sigma_{V}\right)}\boldsymbol{J}_{i}^{*}=2\sqrt{\epsilon}\,\hat{\boldsymbol{J}}^{*}\left(\sigma_{V}\right)\overset{d}{=}2\sqrt{\epsilon N_{c}}\,\hat{\boldsymbol{J}}^{*}\left(\sigma_{K}\right),\label{eq:dfdfdfd}
\end{multline}
introducing the modified temperature 
\begin{equation}
\tilde{\beta}:=2\beta\sqrt{\epsilon N_{c}}=\sqrt{2\log2}\,\beta\,\sqrt{\frac{1-m}{\phi\left(m\right)}}
\end{equation}
and substituting in the formula before we arrive to the expression
\begin{equation}
\langle f_{0}\left[\beta\boldsymbol{\Delta}^{*}\left(\sigma_{V}\right)\right]\rangle_{\boldsymbol{\xi}}\overset{d}{=}\frac{\sum_{\,\alpha\leq2^{K}}\,e^{\,\tilde{\beta}\sqrt{\delta}\,\hat{\boldsymbol{J}}\left(\tau_{K}^{\alpha}\right)\sqrt{K}}f_{0}[\tilde{\beta}\hat{\boldsymbol{J}}^{*}\left(\tau_{K}^{\alpha}\right)]}{\sum_{\alpha\leq2^{K}}e^{\,\tilde{\beta}\sqrt{\delta}\,\hat{\boldsymbol{J}}\left(\tau_{K}^{\alpha}\right)\sqrt{K}}},
\end{equation}
now the REM average formula of Eq. (\ref{eq:REMMMM}) can be applied
in straight fashion, and we see that, at least in the limit of zero
temperature 
\begin{equation}
\langle f_{0}\left[\beta\boldsymbol{\Delta}^{*}\left(\sigma_{V}\right)\right]\rangle_{\boldsymbol{\xi}}^{\lambda}\overset{d}{=}K_{0}^{\lambda}\,\int_{x\in\mathbb{R}}\frac{dx}{\sqrt{2\pi}}\,e^{-\frac{x^{2}}{2}}f_{0}(\tilde{\beta}x)^{\lambda},\label{eq:the result is}
\end{equation}
in this limit the rate parameter $\lambda$ is given by
\begin{equation}
\lambda=\frac{1}{\beta\sqrt{2\delta}}\,\sqrt{\frac{\phi\left(m\right)}{1-m}},
\end{equation}
expanding $\phi\left(m\right)$ for $\beta\rightarrow1$ we find that
near zero temperature
\begin{equation}
\tilde{\beta}\simeq\sqrt{2\log2}\,\sqrt{\frac{\beta}{\psi}},\ \ \lambda\simeq\sqrt{\frac{\psi}{2\beta\delta}}.
\end{equation}
It is an interesting fact that the ratio $N_{c}\simeq e^{2\beta\psi}$
between the original number of spins and the number $K$ at which
the REM average is computed is exponentially diverging in $\beta$:
this suggests a relation with the transposed measure of Section \ref{sec:Kernel-representation},
that can be related to the replicated system, and we interpret as
the measure that is actually used in the 1RSB ansatz with replicas.
\end{proof}
We have found that, at low temperatures, the layer con\-ver\-ges
asymp\-to\-ti\-cal\-ly to a REM, and can be computed using the
properties of PPP. This is not new, in fact, a result similar to Lemma
\ref{lem:(Random-Energy-Model-1} has been obtained for the NPP in
\cite{BCP theorem,BCP 2,Bauke Frnz Mertens}. 

We remark that Eq. (\ref{eq:the result is}) can be extended to any
temperature by noticing that if holds at zero and infinite temperature,
then must hold at any intermediate temperature, alt\-hough with different
coefficients. In fact, it is possible to show that for $\beta\geq\beta'$
holds 
\begin{equation}
E\langle f_{0}\left[\beta\boldsymbol{\Delta}^{*}\left(\sigma_{V}\right)\right]\rangle_{\xi\left(\beta\right)}\leq E\langle f_{0}\left[\beta'\boldsymbol{\Delta}^{*}\left(\sigma_{V}\right)\right]\rangle_{\xi\left(\beta'\right)}
\end{equation}
due to the fact that the variance of $\beta\boldsymbol{\Delta}^{*}\left(\sigma_{V}\right)$
decreases in $m\left(\beta\right)$. Also, when $\tilde{\beta}\geq\tilde{\beta}'$
holds $\lambda(\tilde{\beta})\leq\lambda(\tilde{\beta}')$, then by
Jensen inequality 
\begin{equation}
E\langle f_{0}[\tilde{\beta}\boldsymbol{\Delta}^{*}\left(\sigma_{V}\right)]^{\lambda\left(\tilde{\beta}\right)}\rangle_{\nu}^{1/\lambda\left(\tilde{\beta}\right)}\leq E\langle f_{0}[\tilde{\beta}'\boldsymbol{\Delta}^{*}\left(\sigma_{V}\right)]^{\lambda\left(\tilde{\beta}'\right)}\rangle_{\nu}^{1/\lambda\left(\tilde{\beta}'\right)}.
\end{equation}
Then, notice that in the high temperature limit obviously holds
\begin{equation}
\lim_{\beta\rightarrow0}E\langle f_{0}\left[\beta\boldsymbol{\Delta}^{*}\left(\sigma_{V}\right)\right]\rangle_{\xi\left(\beta\right)}=\lim_{\tilde{\beta}\rightarrow0}E\langle f_{0}[\tilde{\beta}\boldsymbol{\Delta}^{*}\left(\sigma_{V}\right)]^{\lambda\left(\tilde{\beta}\right)}\rangle_{\nu}^{1/\lambda\left(\tilde{\beta}\right)},
\end{equation}
while from Lemma \ref{lem:Let--be_rem} we have found that
\begin{equation}
\lim_{\beta\rightarrow\infty}E\langle f_{0}\left[\beta\boldsymbol{\Delta}^{*}\left(\sigma_{V}\right)\right]\rangle_{\xi\left(\beta\right)}=\lim_{\tilde{\beta}\rightarrow\infty}E\langle f_{0}[\tilde{\beta}\boldsymbol{\Delta}^{*}\left(\sigma_{V}\right)]^{\lambda\left(\tilde{\beta}\right)}\rangle_{\nu}^{1/\lambda\left(\tilde{\beta}\right)}.
\end{equation}
Then, in the TL there must be a mapping between $\beta$ and $\tilde{\beta}$
for any $\beta$, although may be different from that in the proof
of Lemma \ref{lem:Let--be_rem} above, that only holds in the low
temperature limit.

\begin{figure}[p]
\raggedright{}\includegraphics[scale=0.195]{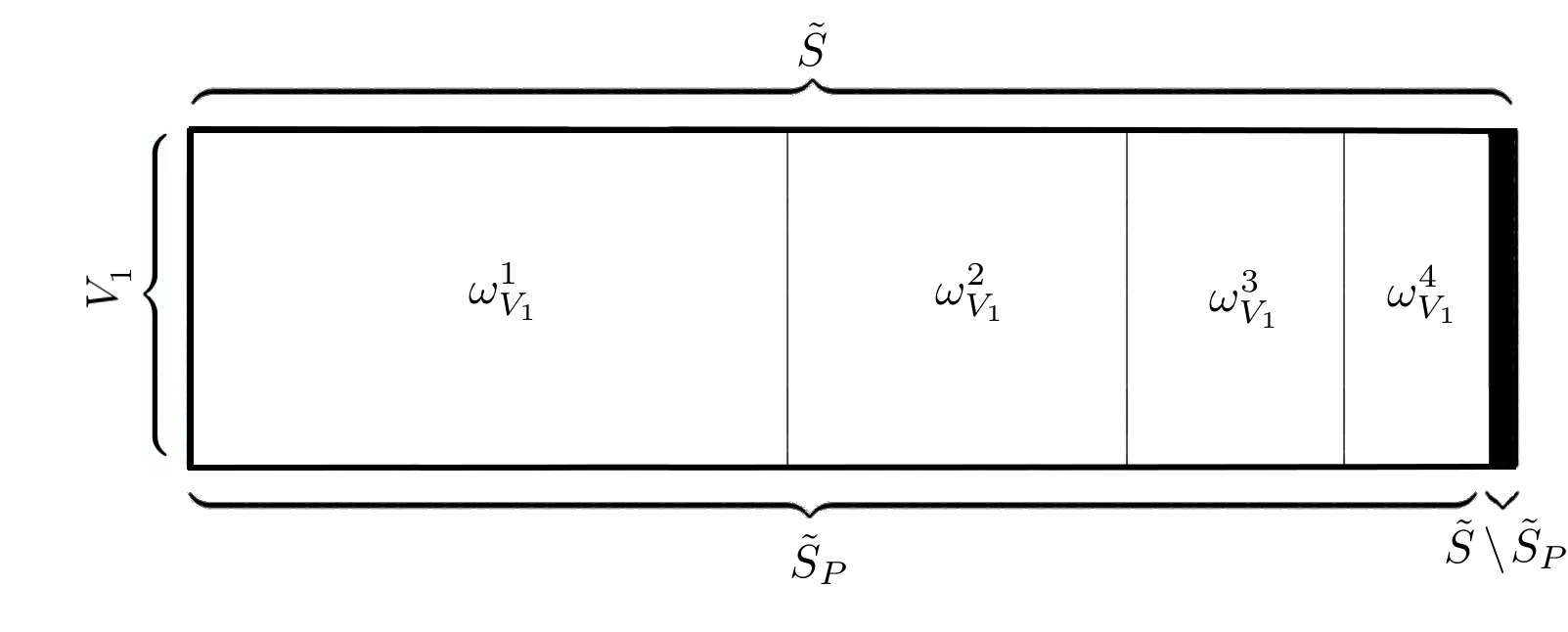}\caption{\label{figANSAZ-1} Sub-kernel from of the first layer $V_{1}$ at
a low temperature and for some fi\-xed realization of the noise.
The support has been relabeled according to a new in\-dex (dependent
from the measure) that orders the states by their probability mass
and where some states have been been removed (call \textit{purified}
index). For some positive $\epsilon$ we select $S_{P}\subset\Omega\left(\left\lfloor m_{1}N_{1}\right\rfloor ,N_{1}\right)$
as the subset of mass $|\tilde{S}_{P}|\geq1-\epsilon$ with the smallest
cardinality, then, the purified index run from $p=1$, the state with
largest weight, to $p=|S_{P}|$, the last before the truncation. We
call $\Omega\left(\left\lfloor m_{1}N_{1}\right\rfloor ,N_{1}\right)\setminus S_{P}$
the irregular set. The ansatz predicts that the states $S_{P}$ on
which $\mu$ concentrates most are randomly selected from $\Omega\left(\left\lfloor m_{1}N_{1}\right\rfloor ,N_{1}\right)$,
which means that the overlap between two states $p$ and $t$ is concentrated
on a deterministic value $q\left(m_{1}\right)$ for all pairs $p\protect\neq t$.
We expect that in the low temperature phase only few states, eventually
only one, will carry most of the probability mass. Notice that for
low temperature the the states overlap almost perfectly with the master
direction, except for some small subset of spins that get flipped.
The localization of such subset is different for each state.}
 
\end{figure}

\begin{figure}
\begin{raggedright}
\includegraphics[scale=0.195]{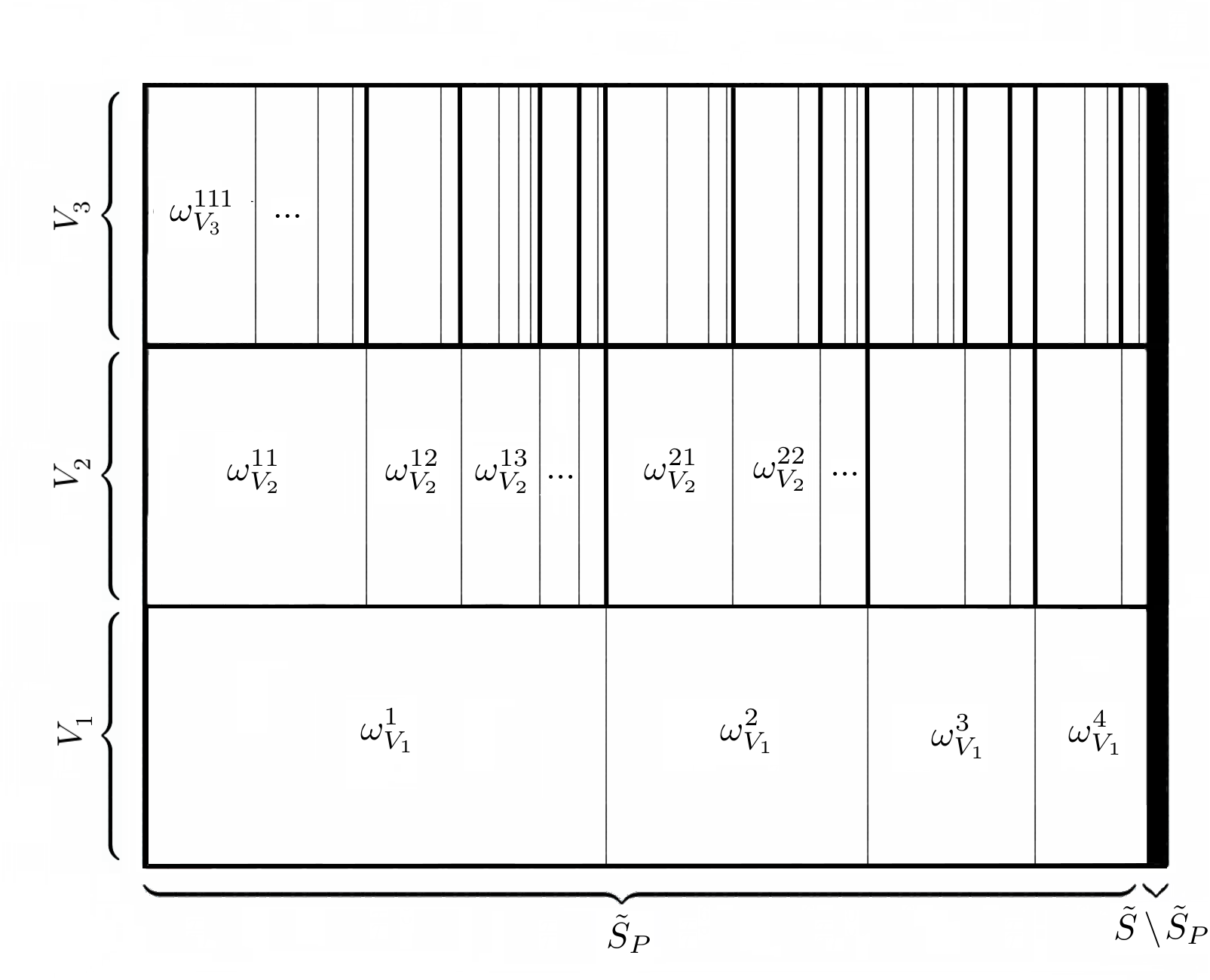}
\par\end{raggedright}

\caption{\label{figANSAZ-2}Kernel representation of the RSB ansatz with $L=3$
for fixed noise. Also in this case we call $S_{P}$ the set of states
that carries most of the probability mass for large $N$, $L$ and
$\beta$, and the states have been relabeled according to a purified
index $p_{1}...\,p_{L}$ where the states of $S\setminus S_{P}$ have
been removed. The ansatz predicts that the layer states are independently
extracted from $\Omega\left(\left\lfloor m_{\ell}N_{\ell}\right\rfloor ,N_{\ell}\right)$
for any $p_{1}...\,p_{\ell-1}\protect\neq t_{1}...\,t_{\ell-1}$.
It is easy to realize that this implies an ultrametric organization
of the overlaps. In fact, consider the states $p_{1}...\,p_{\ell\,}t_{\ell+1}...t_{L}$
and $p_{1}...\,p_{\ell\,}r_{\ell+1}...r_{L}$ of $S_{P}$ with $t_{\ell+1}...\,t_{L}\protect\neq r_{\ell+1}...\,r_{L}$,
the spins of $Q_{\ell}$ will overlap perfectly, while those restricted
to $Q\setminus Q_{\ell}$ will concentrate on some deterministic overlap
that is smaller than one. }
\end{figure}

Lemma \ref{lem:(Random-Energy-Model-1} completes the mandatory tools
to obtain the lower bound of the Pa\-ri\-si for\-mula, using the
Cavity Method in the version of \cite{ASS}, but notice that in our
reasoning we still did not addressed the distribution of the master
direction $\boldsymbol{\omega}_{V}$ itself, that is in fact not strictly
necessary to compute the free en\-er\-gy by Cavity Method. We propose
a conjecture for the full kernel in Figures \ref{figANSAZ-1} and
\ref{figANSAZ-2} but a detailed argument will be given elsewhere.
 
\pagebreak{}

\section{ROSt variables and Parisi functional \label{sec:ROSt-variables-and}}

In this end section we will apply our previous findings and methods
to the cavity re\-pre\-sen\-ta\-tion of the SK incremental pressure
(see for example \cite{ASS,Bolt,Franchini}) and show that by Lemma
\ref{lem:(Random-Energy-Model-1} it can be rewritten to match the
functional of the Parisi formula for the SK model (Parisi functional).
First, we need to represent the the (random) \textit{pressure} 
\begin{equation}
\boldsymbol{p}:=\lim_{N\rightarrow\infty}\frac{1}{N}\log\boldsymbol{Z}
\end{equation}
in terms of some (tractable) functional of $\boldsymbol{\mu}$, this
can be done by cavity method. Fol\-lo\-wing the Random Overlap Structure
(ROSt) oriented derivation of \cite{ASS,Bolt} we define the ROSt
cavity variables, ie the cavity field and the correction term respectively
\begin{equation}
\tilde{\boldsymbol{x}}\left(\sigma_{V}\right)=\frac{1}{\sqrt{N}}\sum_{i}\tilde{\boldsymbol{J}}_{ii}\sigma_{i},\label{eq:cav1}
\end{equation}
\begin{equation}
\tilde{\boldsymbol{y}}\left(\sigma_{V}\right)=\frac{1}{N}\sum_{i<j}\sigma_{i}\tilde{\boldsymbol{J}}_{ij}\sigma_{j}=\frac{1}{\sqrt{N}}\tilde{\boldsymbol{H}}_{sk}\left(\sigma_{V}\right)\overset{d}{=}\frac{1}{\sqrt{N}}\boldsymbol{H}_{sk}\left(\sigma_{V}\right),\label{eq:cav2}
\end{equation}
the last proportional to the Hamiltonian in distribution. Notice that
the above variables are obtained from a noise matrix $\tilde{\boldsymbol{J}}$
that is independent from $\boldsymbol{J}$. We keep a tilde on those
va\-ria\-bles that depends on the new noise. 

Apart from vanishing finite size corrections the Cavity representation
of the in\-cre\-men\-tal pressure in the version of Aizenmann et
al. is \cite{Panchenko,UltraBolt,ASS,Bolt} 
\begin{lem}
(Incremental pressure) The pressure of the Sherrington-Kirckpatrick
model is equal in distribution to the limit 
\begin{equation}
\boldsymbol{p}\stackrel{d}{=}\lim_{N\rightarrow\infty}\,\boldsymbol{A}\left(\tilde{\boldsymbol{x}},\tilde{\boldsymbol{y}},\boldsymbol{\mu}\right),
\end{equation}
where $\boldsymbol{\mu}$ is the SK Gibbs measure, $\tilde{\boldsymbol{x}}$
and $\tilde{\boldsymbol{y}}$ are defined in Eq.s (\ref{eq:cav1}),
(\ref{eq:cav2}) and the functional is 
\begin{multline}
\boldsymbol{A}\left(\tilde{\boldsymbol{x}},\tilde{\boldsymbol{y}},\boldsymbol{\mu}\right)=\log2+\log\sum_{\sigma_{V}\in\Omega^{V}}\boldsymbol{\mu}\left(\sigma_{V}\right)\cosh\left[\beta\,\tilde{\boldsymbol{x}}\left(\sigma_{V}\right)\right]+\\
-\log\sum_{\sigma_{V}\in\Omega^{V}}\boldsymbol{\mu}\left(\sigma_{V}\right)\exp\left[\beta\,\tilde{\boldsymbol{y}}\left(\sigma_{V}\right)\right].\label{eq:ASS-2}
\end{multline}
\end{lem}
\begin{proof}
The result is well known \cite{Parisi-2} and there are multiple routes,
here we follow the derivation that can be found in the last two pages
of \cite{UltraBolt}, originally due to Aizenmann et al. \cite{ASS},
see also \cite{Panchenko,Bolt}. The idea is to relate the partition
function of an $N-$spin system with that of a larger $(N+1)-$system,
then compute the difference between the logarithms of the partition
functions. The key ingredient is the Gaussian sum rule 
\begin{equation}
\boldsymbol{J}_{ij}\sqrt{a+b}\stackrel{d}{=}\boldsymbol{J}_{ij}\sqrt{a}+\tilde{\boldsymbol{J}}_{ij}\sqrt{b},
\end{equation}
where $\tilde{\boldsymbol{J}}$ is a new noise matrix independent
from the $\boldsymbol{J}$. Applying this to the Hamiltonian gives
the following relation that holds in distribution
\begin{multline}
\boldsymbol{H}_{sk}{\textstyle \left(\sigma_{V}\right)}=\frac{1}{\sqrt{N}}\sum_{1\leq i<j\leq N}\sigma_{i}\boldsymbol{J}_{ij}\sigma_{j}\stackrel{d}{=}\\
\stackrel{d}{=}\frac{1}{\sqrt{N+1}}\sum_{1\leq i<j\leq N}\sigma_{i}\boldsymbol{J}_{ij}\sigma_{j}+\frac{1}{\sqrt{N\left(N+1\right)}}\sum_{1\leq i<j\leq N}\sigma_{i}\tilde{\boldsymbol{J}}_{ij}\sigma_{j}.\label{eq:srergerg-1}
\end{multline}
We applied the Gaussian trick before to isolate the correction term,
the partition function can be written as
\begin{equation}
\boldsymbol{Z}\stackrel{d}{=}\sum_{\sigma_{V}\in\Omega^{V}}\exp{\textstyle \left[\beta\sqrt{\frac{N}{N+1}}\,\tilde{\boldsymbol{y}}\left(\sigma_{V}\right)\right]}e^{-\beta\sqrt{\frac{N}{N+1}}\boldsymbol{H}_{sk}{\textstyle \left(\sigma_{V}\right)}}.
\end{equation}
Now consider the system of $N+1$ spins, isolating the last spin gives
\begin{multline}
\boldsymbol{H}_{sk}{\textstyle \left(\sigma_{V\cup\,\left\{ N+1\right\} }\right)}=\frac{1}{\sqrt{N+1}}\sum_{1\leq i<j\leq N+1}\sigma_{i}\boldsymbol{J}_{ij}\sigma_{j}=\\
=\frac{1}{\sqrt{N+1}}\sum_{1\leq i<j\leq N}\sigma_{i}\boldsymbol{J}_{ij}\sigma_{j}+\frac{1}{\sqrt{N+1}}\sigma_{N+1}\sum_{1\leq i\leq N}\boldsymbol{J}_{i,N+1}\sigma_{i}.\label{eq:srergerg}
\end{multline}
Since the sequence $\boldsymbol{J}_{i,N+1}$ is independent from the
other $\boldsymbol{J}$ entries, we can write a more pleasant formula
by using the diagonal terms of $\tilde{\boldsymbol{J}}$ on behalf,
ie we take $\boldsymbol{J}_{i,N+1}=\tilde{\boldsymbol{J}}_{ii}$ so
that the noise relative to the vertex $N+1$ is all expressed in terms
of the $\tilde{\boldsymbol{J}}$ matrix. The associated partition
function is
\begin{equation}
\boldsymbol{Z}^{+}=\sum_{\sigma_{V}\in\Omega^{V}}2\cosh\left[{\textstyle \beta\sqrt{\frac{N}{N+1}}\,\tilde{\boldsymbol{x}}\left(\sigma_{V}\right)}\right]e^{-\beta\sqrt{\frac{N}{N+1}}\boldsymbol{H}_{sk}{\textstyle \left(\sigma_{V}\right)}},
\end{equation}
we have written everything in terms of averages respect to a $N-$system
at slightly shifted temperature. Calling its partition function with
\begin{equation}
\boldsymbol{Z}^{*}=\sum_{\sigma_{V}\in\Omega^{V}}e^{-\beta\sqrt{\frac{N}{N+1}}\boldsymbol{H}_{sk}{\textstyle \left(\sigma_{V}\right)}}
\end{equation}
and dividing by this quantity both $\boldsymbol{Z}^{+}$ and $\boldsymbol{Z}$
we arrive to the expression given in the statement by taking $A\left(\tilde{\boldsymbol{x}},\tilde{\boldsymbol{y}},\boldsymbol{\mu}\right)=\log(\boldsymbol{Z}^{+}/\boldsymbol{Z}^{*})-\log(\boldsymbol{Z}/\boldsymbol{Z}^{*})$,
apart from a rescaling of the temperature that becomes negligible
in the TL. Notice that since this expression is a representation for
the incremental free energy and not the actual free energy, then the
proper relation with $\boldsymbol{p}$ would be rather 
\begin{equation}
\boldsymbol{p}\geq\liminf_{N\rightarrow\infty}\,\boldsymbol{A}\left(\tilde{\boldsymbol{x}},\tilde{\boldsymbol{y}},\boldsymbol{\mu}\right),
\end{equation}
but for the SK model it can be shown that the bound is tight \cite{Panchenko,Bolt}.
\end{proof}
\noindent The cavity formula is then written using Lemma \ref{11(Filtration-of-).}
as
\begin{multline}
\boldsymbol{A}\left(\tilde{\boldsymbol{x}},\tilde{\boldsymbol{y}},\boldsymbol{\xi}\right)=\log2+\log\,\langle\,...\,\langle\,\cosh\left[\beta\,\tilde{\boldsymbol{x}}\left(\sigma_{Q_{L}}\right)\right]\,\rangle_{\boldsymbol{\xi}_{L}}...\,\rangle_{\boldsymbol{\xi}_{1}}+\\
-\log\,\langle\,...\,\langle\,\exp\left[\beta\,\tilde{\boldsymbol{y}}\left(\sigma_{Q_{L}}\right)\right]\,\rangle_{\boldsymbol{\xi}_{L}}...\,\rangle_{\boldsymbol{\xi}_{1}}.\label{eq:ASS}
\end{multline}
Here is an important step. Let rewrite the formula once again according
to the partition $\mathcal{V}$ by introducing the variable
\begin{equation}
\tilde{\boldsymbol{z}}_{\ell}\left(\sigma_{V_{\ell}}\right)\sqrt{\left|V_{\ell}\right|}:=\sum_{i\in V_{\ell}}\tilde{\boldsymbol{J}}_{ii}\sigma_{i},
\end{equation}
that is the $V_{\ell}$ component of the cavity field normalized by
the square root of the number of spins $\left|V_{\ell}\right|$, and
the variable
\begin{equation}
\tilde{\boldsymbol{g}}_{\ell}\left(\sigma_{Q_{\ell}}\right)\sqrt{\left|W_{\ell}\right|}:=\sum_{\left(i,j\right)\in W_{\ell}}\sigma_{i}\tilde{\boldsymbol{J}}_{ij}\sigma_{j},
\end{equation}
that is the $W_{\ell}$ component of the correction term, this normalized
with the square root of the number of terms $|W_{\ell}|$ contributing
to $\tilde{\boldsymbol{H}}_{\ell}\left(\sigma_{Q_{\ell}}\right)$
of Lemma \ref{lem:(Layer-States-of}. 

Notice that for fixed $\sigma_{Q_{\ell}}$ both variables are normally
distributed. We rewrite the cavity variables in terms of the previous,
the first is as follows
\begin{equation}
\tilde{\boldsymbol{x}}\left(\sigma_{V}\right)\sqrt{N}=\sum_{\ell\leq L}\tilde{\boldsymbol{z}}_{\ell}\left(\sigma_{V_{\ell}}\right)\sqrt{\left|V_{\ell}\right|}
\end{equation}
while for the correction term we can follow the same decomposition
presented in Lemma \ref{lem:(Layer-States-of} and find
\begin{equation}
\tilde{\boldsymbol{y}}\left(\sigma_{V}\right)\sqrt{2N}=\sum_{\ell\leq L}\tilde{\boldsymbol{g}}_{\ell}\left(\sigma_{Q_{\ell}}\right)\sqrt{\left|W_{\ell}\right|}
\end{equation}
where $1/\sqrt{2}$ (and not $1/2$) comes from removing the $i<j$
constraint under the assumption that $\tilde{\boldsymbol{J}}$ is
asymmetric almost surely. Recall that by definition the sizes of the
sets are
\begin{equation}
\left|V_{\ell}\right|=\left|Q_{\ell}\right|-\left|Q_{\ell-1}\right|=\left(q_{\ell}-q_{\ell-1}\right)N,
\end{equation}
\begin{equation}
\left|W_{\ell}\right|=\left|Q_{\ell}\right|^{2}-\left|Q_{\ell-1}\right|^{2}=\left(q_{\ell}^{2}-q_{\ell-1}^{2}\right)N^{2}.
\end{equation}

{\tiny{}~}{\tiny \par}

\noindent We can already recognize two familiar coefficients, in particular,
these relations allow to identify the sizes of the sets $|Q_{\ell}|$
with the overlap parameters $q_{\ell}$ as usually intended in the
RSB theory. Then, substituting these expressions into the cavity formula
before we arrive to 
\begin{multline}
\boldsymbol{A}\left(q,\tilde{\boldsymbol{z}},\tilde{\boldsymbol{g}},\boldsymbol{\xi}\right)=\log\,\langle\,...\,\langle2\cosh{\textstyle \left[\beta\,{\textstyle \sum_{\ell}}\,\tilde{\boldsymbol{z}}_{\ell}\left(\sigma_{V_{\ell}}\right)\sqrt{q_{\ell}-q_{\ell-1}}\,\right]}\rangle_{\boldsymbol{\xi}_{L}}...\,\rangle_{\boldsymbol{\xi}_{1}}+\\
-\log\,\langle\,...\,\langle\,\exp{\textstyle \left[{\textstyle \frac{\beta}{\sqrt{2}}\sum_{\ell}}\,\tilde{\boldsymbol{g}}_{\ell}\left(\sigma_{Q_{\ell}}\right)\sqrt{q_{\ell}^{2}-q_{\ell-1}^{2}}\,\right]}\rangle_{\boldsymbol{\xi}_{L}}...\,\rangle_{\boldsymbol{\xi}_{1}}.\label{eq:ASS-1}
\end{multline}

As one can easily see, the Definition \ref{11(Filtration-of-).} provides
a natural description of the ROSt probability space and its variables.
Notice that up to this point all the things that we did on $\boldsymbol{A}$
depend on the partition of $V$ and hold in general, nonetheless,
both the cavity field and the correction term (called \textit{fugacity
variable} in \cite{ASS}) are now expressed using a common kernel
base. In the following theorem, we show how to obtain the functional
that appear in the celebrated variational formula by Parisi. Concerning
the version of the functional, we refer to the one given in reference
\cite{Bolt} for a comparison. 
\begin{thm}
(Parisi functional) Lemma \ref{lem:(Random-Energy-Model-1} applied
to the cavity representation of Eq. (\ref{eq:ASS-1}) gives the Parisi
functional 
\begin{equation}
A_{P}\left(q,\lambda\right)=\log2+\log Y_{0}-\frac{\beta^{2}}{4}\sum_{\ell}\lambda_{\ell}\left(q_{\ell}^{2}-q_{\ell-1}^{2}\right),
\end{equation}
where $Y_{0}$ given by the recursion $\boldsymbol{Y}_{\ell-1}^{\lambda_{\ell}}=E_{\ell}\boldsymbol{Y}_{\ell}^{\lambda_{\ell}}$
applied to the initial condition 
\begin{equation}
\boldsymbol{Y}_{L+1}=\cosh\left(\beta{\textstyle \sum_{\ell}}\,\tilde{\boldsymbol{z}}_{\ell}\sqrt{q_{\ell}-q_{\ell-1}}\right),
\end{equation}
with $\tilde{\boldsymbol{z}}_{\ell}$ i.i.d. normally distributed
and $E_{\ell}$ normal average acting on $\tilde{\boldsymbol{z}}_{\ell}$.\end{thm}
\begin{proof}
From Lemma \ref{lem:(Random-Energy-Model-1} applied to Eq. (\ref{eq:ASS-1})
we find 
\begin{equation}
{\textstyle \sum_{\sigma_{V_{\ell}\in\Omega^{V_{\ell}}}}}\boldsymbol{\xi}_{\ell}\left(\sigma_{Q_{\ell}}\right)\boldsymbol{f}\left[\tilde{\boldsymbol{z}}_{\ell}\left(\sigma_{V_{\ell}}\right)\right]\stackrel{d}{=}K_{\ell}\,\langle\,\boldsymbol{f}\left[\tilde{\boldsymbol{z}}_{\ell}\left(\boldsymbol{\sigma}_{V_{\ell}}\right)\right]{}^{\lambda_{\ell}}\rangle_{\nu}^{1/\lambda_{\ell}}\label{eq:6.22}
\end{equation}
in distribution for some constant $K_{\ell}$, and the same can be
done for $\tilde{\boldsymbol{g}}_{\ell}$, since in the Eq. (\ref{eq:6.22})
the uniform average $\nu$ is applied, we can safely take
\begin{equation}
\tilde{\boldsymbol{z}}_{\ell}\left(\sigma_{V_{\ell}}\right)\stackrel{d}{=}\tilde{\boldsymbol{z}}_{\ell},\ \tilde{\boldsymbol{g}}\left(\sigma_{Q_{\ell}}\right)\stackrel{d}{=}\tilde{\boldsymbol{g}_{\ell}}
\end{equation}
with $\tilde{\boldsymbol{z}}_{\ell}$ and $\tilde{\boldsymbol{g}}_{\ell}$
independent and normally distributed for all $\ell$, then
\begin{equation}
\langle\,f\left[\tilde{\boldsymbol{z}}_{\ell}\left(\boldsymbol{\sigma}_{V_{\ell}}\right)\right]{}^{\lambda_{\ell}}\rangle_{\nu}^{1/\lambda_{\ell}}\stackrel{d}{=}{\textstyle [E_{\ell}f\left(\tilde{\boldsymbol{z}}_{\ell}\right)^{\lambda_{\ell}}]^{1/\lambda_{\ell}}},
\end{equation}
and do the same for $\tilde{\boldsymbol{g}}\left(\sigma_{Q_{\ell}}\right)$.
Now, start from the initial condition $\boldsymbol{Y}_{L+1}$ and
apply the averages down to $\ell=0$. We arrive at
\begin{equation}
\langle\,...\,\langle\,\cosh\left[\beta{\textstyle \sum_{\ell}}\,\tilde{\boldsymbol{z}}_{\ell}\left(\boldsymbol{\sigma}_{V_{\ell}}\right)\sqrt{q_{\ell}-q_{\ell-1}}\,\right]\rangle_{\boldsymbol{\xi}_{L}}...\,\rangle_{\boldsymbol{\xi}_{1}}\stackrel{d}{=}Y_{0}\exp\left({\textstyle \sum_{\ell}}\log K_{\ell}\right).
\end{equation}
Then we can compute the correction term in the same way, finding 
\begin{multline}
\langle\,...\,\langle\,\exp\left[{\textstyle \frac{\beta}{\sqrt{2}}\sum_{\ell}}\,\tilde{\boldsymbol{g}}\left(\boldsymbol{\sigma}_{Q_{\ell}}\right)\sqrt{q_{\ell}^{2}-q_{\ell-1}^{2}}\right]\rangle_{\boldsymbol{\xi}_{L}}...\,\rangle_{\boldsymbol{\xi}_{1}}\stackrel{d}{=}\\
\stackrel{d}{=}\exp\left[{\textstyle \frac{\beta^{2}}{4}\sum_{\ell}}\lambda_{\ell}\left(q_{\ell}^{2}-q_{\ell-1}^{2}\right)+{\textstyle \sum_{\ell}}\log K_{\ell}\right].\label{eq:www}
\end{multline}
Putting together, the contributions depending from $K_{\ell}$ cancel
out, 
\begin{equation}
\boldsymbol{A}\left(q,\tilde{\boldsymbol{z}},\tilde{\boldsymbol{g}},\boldsymbol{\xi}\right)\stackrel{d}{=}A_{P}\left(q,\lambda\right),\label{eq:final}
\end{equation}
and we obtained the Parisi functional as is presented in \cite{Bolt}.
\end{proof}
At this point we can easily understand also the origin of the functional
parameters appearing in the Parisi formula. The variables $q_{\ell}$
control the energy contributions due to the new spin (actually is
the absence of it) and are determined by the relative sizes of the
sets in the partition $\mathcal{V}$, while the lambda parameters
$\lambda_{\ell}$ control the cascade of Point Processes. 

A fundamental aspect of the Parisi ansatz, which we do not address
here, is the special direction of the variational principle to obtain
the pressure (one takes the inferior limit of the functional instead
of the superior as in Boltzmann theory). Concerning our previous computations,
up to now we assumed $q_{\ell}$ and $\lambda_{\ell}$ fixed to the
correct SK value, but one can immediately write the lower bound by
varying them 
\begin{equation}
E\left(\boldsymbol{p}\right)\geq\inf_{q,\lambda}A_{P}\left(q,\lambda\right).
\end{equation}
Clearly, the hard part is to prove that this inequality is tight.
At least for the SK model this has been obtained by proving a matching
upper bound via Gaussian interpolation techniques (Guerra-Toninelli
interpolation \cite{Guerra}). 

Then, the physical meaning of the Parisi variational formula would
be in some equilibration condition between the original system and
the contributions from the new layer, here composed by just one spin
(but one can add more and find the same result). Moreover, it has
been proven in \cite{AuffingerChen} that the Parisi functional $A_{P}\left(q,\lambda\right)$
has a unique minimizer, indicating that such equilibration process
has only one result. Since the Parisi principle prescribes to maximize
the incremental free energy functional the exact mechanism behind
is still not evident, by the way, assuming that the thermodynamic
limit exists, then starting from Eq.(\ref{eq:ASS-1}) and applying
Jensen inequality 
\begin{equation}
E\left(\boldsymbol{p}\right)\leq\log E\left[\exp\boldsymbol{A}\left(q,\tilde{\boldsymbol{z}},\tilde{\boldsymbol{g}},\boldsymbol{\xi}\right)\right],
\end{equation}
and since by Eq.(\ref{eq:final}) the functional is distributed like
$A_{P}\left(q,\lambda\right)$, that is non-random, one finds that
$A_{P}\left(q,\lambda\right)$ is also an upper bound for the expectation
of the incremental free energy for any value of $q$ and $\lambda$. 

\pagebreak{}

\section{Conclusions and outlooks}

We have described a mathematical formalism that provides a rigorous
framework to handle spin glass problems at finite and infinite volume
and in a constructive way. In particular, the analogue pure states
of Definition \ref{11(Filtration-of-).} allow a constructive approach
to the cavity method with ROSt (Sections \ref{sec:The-Ansatz} and
\ref{sec:ROSt-variables-and}), and eventually provide a scheme to
go from 1RSB to L-RSB once the 1RSB approximation is known. In Section
\ref{sec:The-Ansatz} we give a detailed analysis of the martingale
representation of \cite{Franchini} and its relation to the Random
Energy Model. These manipulations show that a constructive approach
to RSB is at least possible without averages and non-standard algebraic
tricks.

Concerning the extensibility of the method beyond the SK model, we
remark that the arguments of the first three sections and their mathematical
formalism are very general, and allow for much in-depth analysis and
ge\-ne\-ra\-li\-za\-tions. For example, it can be used to deal
with any Hamiltonian
\begin{equation}
H\left(\sigma_{V}\right)=\frac{1}{g\left(H\right)}\,\sum_{i\in V}\sum_{j\in V}\sigma_{i}H_{ij}\sigma_{j},
\end{equation}
with $H_{ij}$ being any interaction matrix with fancy underlying
topology, and ge\-ne\-ra\-li\-zed to bipartite and multi-spin
interactions by considering higher dimensional kernels (multi-kernels).
Also, the kernel formalism allowed to introduce the tran\-spo\-sed
mea\-sure, an interesting pro\-ba\-bi\-lis\-tic object that we
interpret in the proof of Lemma \ref{lem:Let--be_rem} as physically
significant to the replica space. 

The arguments of Sections \ref{sec:The-Ansatz}, \ref{sec:Thermodynamics-of-the}
and \ref{sec:ROSt-variables-and} are also very general, and they
can be di\-rec\-tly applied for noise that is not Gaussian, and
also to the multipartite models. For example, it is possible to repeat
the same analysis for the Little model \cite{BrunettiRitortParisi},
a bipartite system defined by the Hamiltonian
\begin{equation}
\boldsymbol{H}_{Little}\left(\sigma_{V},\tau_{V}\right):=\frac{1}{\sqrt{N}}\sum_{i\in V}\sum_{j\in V}\sigma_{i}\boldsymbol{J}_{ij}\tau_{j}.
\end{equation}
We can write the same layer decomposition of Lemma \ref{lem:(Layer-States-of},
obtaining
\begin{multline}
\boldsymbol{H}_{\ell}\left[\,\sigma_{V_{\ell}},\boldsymbol{h}_{V_{\ell}}\left(\sigma_{Q_{\ell-1}}\right),\tau_{V_{\ell}},\boldsymbol{h}_{V_{\ell}}\left(\tau_{Q_{\ell-1}}\right)\right]=\\
=\sqrt{q_{\ell}-q_{\ell-1}}\,\boldsymbol{H}\left(\sigma_{V_{\ell}},\tau_{V_{\ell}}\right)+\sqrt{q_{\ell-1}}\,\sigma_{V_{\ell}}\cdot\boldsymbol{h}_{V_{\ell}}\left(\tau_{Q_{\ell-1}}\right)+\\
+\sqrt{q_{\ell-1}}\,\tau_{V_{\ell}}\cdot\boldsymbol{h}_{V_{\ell}}\left(\sigma_{Q_{\ell-1}}\right),\label{eq:ssvvv}
\end{multline}
then we apply Lemma \ref{lem:(IO-model)-For}, that establish the
irrelevance of the energy contribution from the core $\boldsymbol{H}\left(\sigma_{V_{\ell}},\tau_{V_{\ell}}\right)$,
and then proceed as in the SK case. 

There is a limitation in that Lemma \ref{lem:(IO-model)-For} is expected
to hold for fully connected mo\-dels only. In fact, in the case of
sparsely connected models the contribution from the layer core to
the energy could be still relevant. By the way, detailed computations
should be possible for the Bethe Lattice and other sparse but mean-field
structured models, in this respect it would be interesting to confront
with the loop expansion method of Chertkov, Chernyak, Xiao and Zhou
\cite{Xiao Zhou}. 

Concerning finite dimensional lattice models, consider for example
the Ising Spin Glass on a $d-$dimensional lattice, the Hamiltonian
is
\begin{equation}
\boldsymbol{H}_{d}\left(\sigma_{V}\right):=\frac{1}{g\left(\Lambda\right)}\sum_{i\in V}\sum_{i<j}\sigma_{i}\boldsymbol{J}_{ij}\Lambda_{ij}\sigma_{j},
\end{equation}
where $\Lambda_{ij}$ is the adjacency matrix of the considered lattice,
and 
\begin{equation}
\boldsymbol{h}_{i}\left(\sigma_{V_{\ell-1}}\right):=\frac{1}{g\left(\Lambda\right)}\sum_{j\in V_{\ell}}\boldsymbol{J}_{ij}\Lambda_{ij}\sigma_{j}
\end{equation}
are cavity fields that describe the energy contributions from the
interface between the spins of $V_{\ell}$ and $V_{\ell-1}$. Then
we can apply the layer decomposition to obtain 
\begin{equation}
\boldsymbol{H}_{\ell}\left[\,\sigma_{V_{\ell}},\boldsymbol{h}_{V_{\ell}}\left(\sigma_{V_{\ell-1}}\right)\right]=\boldsymbol{H}_{d-1}\left(\sigma_{V_{\ell}}\right)+\sum_{j\in V_{\ell}}\sigma_{i}\cdot\boldsymbol{h}_{i}\left(\sigma_{V_{\ell-1}}\right),
\end{equation}
where $\boldsymbol{H}_{d-1}$ is the Hamiltonian of the layer's core,
that is a $d-1$ dimensional mo\-del. In this case the Lemma \ref{lem:(IO-model)-For}
cannot be applied, because the core and the in\-ter\-face con\-tributions
to the layer energy have comparable sizes, but it should be possible
to obtain results with a proper choice of the $Q_{\ell}$ sequence
and the lat\-tice geometry, a short discussion of this can be found
in \cite{Franchini}.

Although we stop here for this paper, we are persuaded that the kernel
framework could provide the ground to systematize many of the known
relations between the spin glass problems and other important fields
of physics and mathematics, including graph theory \cite{Lovasz}
(the Section \ref{sec:Kernel-filtration} provides the connection
with \cite{ACO-1,ACO2,ACO3,ACO4} and \cite{Lovasz}), neural networks
\cite{SommersCrisanti,Battaglia} (notice that the structure proposed
in Section \ref{sec:The-Ansatz} already resembles a layer neural
network), polymers (the layer states of Definition \ref{11(Filtration-of-).}
have be\-en us\-ed already to compute an urn model ap\-proximation
of the Range Problem on fi\-ni\-te dimensional lattices in \cite{Franchini Polymers})
soft granular matter \cite{Sfere dure Benetti}, Nelson mechanics
\cite{Stochastic Mechanics,Planar} (through the commutation properties
expressed by Lemma \ref{lem:(Correlations-and-Overlaps)}), analytic
number theory \cite{BCP theorem,BCP 2,Bauke Frnz Mertens,Number theory,Parisi sourlas,V A Avetisov}
and algorithmic op\-ti\-mi\-za\-tion (see \cite{BPAlgorithms}
for a review). Further in\-ves\-tigations will establish where this
could eventually bring. 

Apart from the spin glass theory and RSB, the kernel representation
provides a for\-mal ground to confront datasets from real experiments
with kernels from spin glass pro\-blems. We remark that datasets
of this kind are already available, for example, from neural activity
measurements in mammals, most interesting are those from single neuron
spike detections (see the kernels obtained in the re\-markable experiments
of Clawson et al. in \cite{Battaglia}).

\pagebreak{}

\section{Acknowledgments}

I wish to thank Amin Coja-Oghlan (Goethe University, Frankfurt) for
explaining the ideas presented in \cite{ACO-1}, and their connection
with Graph Theory and Replica Symmetry Breaking. I also wish to thank
Giorgio Parisi, Fran\-ce\-sco Guerra, Pie\-tro Caputo, Gioia Pica,
Nicola Kistler, Demian Battaglia, Pan Liming, Fran\-ce\-sco Concetti
and Riccardo Balzan for interesting discussions and suggestions. This
project has received funding from the European Research Council (ERC)
under the European Union’s Seventh Framework Programme (grant agreement
No {[}278857{]}) and the European Union’s Horizon 2020 research and
innovation programme (grant agreement No {[}694925{]}).

\end{document}